\documentclass[11pt,a4paper]{article}
\pdfoutput=1

\usepackage{jheppub}
\usepackage{amsfonts, amsmath, bm}
\usepackage{enumitem}
\usepackage{xspace}
\usepackage{xcolor, graphicx, tikz}

\makeatletter
\def\@fpheader{\relax}
\makeatother

\newcommand\comment[1]{}
\newcommand\be{\begin{equation}}
\newcommand\ee{\end{equation}}
\newcommand\bea{\begin{eqnarray}}
\newcommand\eea{\end{eqnarray}}

\newcommand\Comment[1]{}

\newcommand{\CY}{Calabi--Yau\xspace}
\newcommand{\Vol}{\mathrm{Vol}}
\newcommand{\Tr}{\mathrm{Tr}}
\newcommand{\dd}{\mathrm{d}}

\newcommand{\cV}{\mathcal{V}}
\newcommand{\cE}{\mathcal{E}}

\newcommand{\WP}{\mathrm{WP}}
\newcommand\num[1]{{\boldsymbol{\mathsf{#1}}}}

\def\IR{\mathbb{R}}

\title{\centering What to do with a Ricci-flat Calabi--Yau metric?}

\author[*]{Per Berglund,}
\author[\dagger]{Tristan H\"ubsch,}
\author[\ddagger,\S]{Vishnu Jejjala}

\affiliation[*]{Department of Physics and Astronomy, University of New Hampshire, Durham, NH 03824, USA}

\affiliation[\dagger]{Department of Physics and Astronomy, Howard University, Washington, DC 20059, USA}

\affiliation[\ddagger]{Mandelstam Institute for Theoretical Physics, School of Physics, and NITheCS,\\
University of the Witwatersrand, Johannesburg, WITS 2050, South Africa}

\affiliation[\S]{The NSF Institute for Artificial Intelligence and Fundamental Interactions (IAIFI)\\
and Department of Physics, Northeastern University, Boston, MA 02115, USA}

\emailAdd{per.berglund@unh.edu}
\emailAdd{thubsch@howard.edu}
\emailAdd{v.jejjala@wits.ac.za}

\abstract{%
Numerical approximations to Ricci-flat \CY metrics make it possible to move beyond the topological and holomorphic data that have traditionally dominated explicit string compactifications.
This article explains what new physics and mathematics become accessible once the metric, and eventually the associated Hermitian Yang--Mills bundle data, can be computed.
In heterotic compactifications, such data are needed to determine matter K\"ahler metrics, canonically normalized Yukawa couplings, Kaluza--Klein spectra, threshold effects, soft terms, and other non-holomorphic ingredients of the four-dimensional effective action.
More broadly, numerical \CY geometry provides quantitative input for moduli stabilization, $\alpha'$-corrected backgrounds, de~Sitter model building, axion physics, swampland distance tests, and compactifications in which the internal geometry varies over spacetime.
Geometric data permit a computational approach to long-standing mathematical questions involving special Lagrangian submanifolds, SYZ fibrations, mirror symmetry, calibrated geometry, metric degeneration, restrictions of Ricci-flat metrics to fibers, and the search for analytic or semi-analytic structures.
We present these directions as a roadmap for future work.
}

\begin{document}
\maketitle
\flushbottom

\section{Introduction}
\label{s:intro}
\label{s:what} 
Yau's theorem~\cite{Yau:1977ms,Yau:1978rc} guarantees the existence of a unique Ricci-flat K\"ahler metric in each K\"ahler class on a compact, complex \CY manifold, but that metric is not known in closed form in essentially any compact example relevant to string theory.
This inconvenient truth created a structural bottleneck for decades:
One could classify topological and various refined compactification data, compute cohomology, derive selection rules, and sometimes determine holomorphic superpotentials, but virtually never also the canonically normalized quantities that determine the ``actual physics'' --- to be compared with actual experiments.
Numerical approaches based on balanced metrics, generalized Donaldson algorithms, and more recently machine learning have changed the status of the problem, from ``the metric exists'' to ``the metric can be computed with useful, and increasingly quantifiable, a posteriori error diagnostics''~\cite{Headrick:2005ch, Donaldson:2005hvr, Douglas:2006rr, Headrick:2009jz, Ashmore:2019wzb, Anderson:2020hux, Douglas:2020hpv, Jejjala:2020wcc, Douglas:2021zdn, Larfors:2022nep, Berglund:2022gvm, Hendi:2024yin}.

The practical meaning of this development is hard to overstate.
In heterotic superstring compactifications, the internal metric enters the Hodge star, the Laplacian, the harmonic representatives, the matter K\"ahler metrics, the moduli space metrics, the normalized Yukawa couplings, the Kaluza--Klein spectra, and eventually the soft terms.
In type~II and F-theory compactifications, it controls divisor volumes, instanton actions, field space distances, warping source terms, and the canonical normalization needed for precision moduli stabilization and cosmology.
In geometry, the metric is the gateway to calibrated submanifolds, Laplace spectra, special Lagrangians, degeneration limits, and genuinely metric versions of mirror symmetry and the SYZ conjecture.

A useful slogan comes to mind:
\begin{quote}
\emph{Topology tells you what can happen;\\ the metric tells you how much of it actually happens and where.}
\end{quote}
This note is organized around this slogan.
The focus is not on the numerical algorithms for computing the Ricci-flat Calabi--Yau metric~\cite{Qi2020,Larfors:2021pbb,Gerdes:2022nzr,Berglund:2024psp} or potential improvements, but on the research program unlocked once this metric has been computed, and on the resulting directions that subsequently become available.

\subsection{What does a Ricci-flat Calabi--Yau metric give you?}

Let $X$ be a \CY threefold with K\"ahler class $[J]$, a nowhere vanishing holomorphic three-form $\Omega$, and a {\em numerically obtained\/} Ricci-flat metric $\num{g}$ whose K\"ahler form $\omega_{\num{g}}$ represents $[J]$.
From this data one obtains, at least in principle,
\begin{align}
\dd\mathrm{Vol}_{\num{g}} &= \tfrac1{3!}\,\omega_{\num{g}}^3, &
\Delta_{\num{g}} &= \dd\dd^{\dagger} + \dd^{\dagger}\dd, &
*_{\num{g}} &: \wedge^p T^*X \to \wedge^{6-p}T^*X,
\end{align}
as well as the Levi-Civita connection, geodesics, norms of differential forms, curvature tensors and polynomial curvature invariants.
The metric therefore gives direct access to:
\begin{itemize}[topsep=-3pt, itemsep=-1pt]
  \item volumes of curves, divisors, and the whole manifold,
  \item harmonic representatives of cohomology classes,
  \item eigenvalues and eigenfunctions of scalar, form, and Dirac Laplacians,
  \item Green's operators and propagators on the compact space,
  \item pointwise curvature invariants and local measures of the validity of the derivative expansion.
\end{itemize}
Furthermore, if a family $\num{g}(m,\bar m)$ is known over the moduli space rather than only at one point therein, then one also gets moduli space metrics, geodesic distances in the moduli space, and moduli dependence of masses and (physically normalized!) couplings.

For heterotic compactifications, this is necessary but, for realistic models, not sufficient.
By the Donaldson--Uhlenbeck--Yau theorem, a holomorphic polystable vector bundle $V\to X$ of vanishing slope in the relevant K\"ahler class admits a Hermitian Yang--Mills (HYM) connection $A$~\cite{Donaldson:1985zz,Uhlenbeck:1986de}; in practice one needs a {\em numerical approximation\/} $\num{A}$ as well.
The zero modes are bundle-valued forms that are harmonic with respect to the coupled Dolbeault or Dirac operator, so the full pipeline is really
\begin{equation}
\begin{aligned}
(X,\num{g})~~ \leadsto&~~ (V,\num{A})~~ \leadsto~~ \text{$\bar\partial_{\num{A}}$-harmonic bundle-valued modes},\nonumber\\
&~~\leadsto~~ \text{4d effective action, target space physics}.
\end{aligned}
\end{equation}
A similar remark applies in type~II/F-theory compactifications: the metric alone is not the compactification; one also needs orientifold data, fluxes, branes, and often warping.

Thus, the correct viewpoint is that a numerical \CY metric provides a \emph{platform}.
It supplies the differential geometric infrastructure on top of which compactification physics can be computed. This roadmap gives guidance for where the metric is useful.
Some of the problems concern established formul\ae\ that require the metric as input.
There are near-term numerical tasks for which current methods plausibly provide the needed data.
Finally, in more speculative directions, the metric supplies a starting point rather than a complete solution.

\subsection{Improvements from point selection}
Numerical approximations are sensitive to point sampling.
The standard point selection strategy for Calabi--Yau hypersurfaces and complete intersections is to sample not the variety directly, but auxiliary linear data in the ambient projective space.
For a hypersurface $X=\{p=0\}\subset \mathbb P^n$, one draws random vectors $v\in S^{2n+1}\subset \mathbb C^{n+1}$, quotients by the Hopf $S^1$-action to obtain Fubini--Study distributed points in $\mathbb P^n$, builds random projective lines
\begin{equation}
L_{ij}=\{[v_i+\lambda v_j]\mid \lambda\in\mathbb C\}\subset\mathbb P^n ~,
\end{equation}
and then takes $L_{ij}\cap X$ as the point sample.
The reason this construction is mathematically controlled is the Shiffman--Zelditch equidistribution theorem: zeros of random holomorphic sections of a positive Hermitian line bundle are distributed, in the large-degree limit, with respect to the curvature/Fubini--Study measure.
Thus the sampler is uniform in the $SU(n+1)$-invariant Fubini--Study sense, and it gives a known measure against which the Monge--Amp\`ere and consistency losses may be estimated.
This is the sampling philosophy used in the earlier numerical metric literature and inherited by many later machine learning implementations~\cite{Shiffman:1998,Braun:2007sn,Anderson:2010ke,Ashmore:2019wzb}.

Bertrand's paradox is the elementary warning that geometric randomness is not intrinsic until the measure is specified.
A chord of a circle may be chosen by two random endpoints on the circumference, by choosing a random radius and then a random perpendicular, or by choosing a random midpoint in the disk; each prescription looks natural, but the probability that the chord is longer than a side of the inscribed equilateral triangle is, respectively, $1/3$, $1/2$, or $1/4$~\cite{Bertrand:1889}.
Jaynes' reformulation emphasizes that the missing datum is an invariance principle: if the size and location of the circle are not specified, one should demand scale and translation invariance~\cite{Jaynes:1973}.
The analogy with Calabi--Yau point selection is direct.
Saying that points are ``uniform'' on $X$ is not meaningful until one says with respect to which volume form --- the ambient Fubini--Study volume form, the holomorphic volume form, a known flat metric in a special example, or the unknown Ricci-flat metric being learned.
Different choices can lead to different quadrature errors and hence to different apparent convergence for the same neural network.

The elliptic curve provides a clean diagnostic because the intrinsic flat metric is known.
If $T^2$ is realized as the Fermat cubic in $\mathbb P^2$, one can compare two point clouds: one obtained from the Shiffman--Zelditch/Fubini--Study line construction, and one obtained by sampling the flat fundamental domain of $\mathbb C/\Lambda$ and mapping to the cubic.
Using the same neural network model, namely the spectral network architecture employed in~\cite{Berglund:2022gvm}, the flat metric sampling gives a two order of magnitude improvement in the loss function after training for the same number of epochs compared to the Shiffman--Zelditch sampling~\cite{Berglund:unpubPointSelection}.
This is consistent with the broader numerical lesson that spectra and metric dependent observables are sensitive to the quality and distribution of the point cloud~\cite{Ahmed:2023cnw}.
The conclusion is not that the Fubini--Study sampler is wrong, but rather that it is an ambient, not intrinsic, notion of randomness; when the intrinsic measure is known, sampling with respect to it can reduce the variance and bias of the empirical Monge--Amp\`ere loss.

An improved sampling strategy suggested by~\cite{Ruehle:Pollica2025,Lust:2026mys} is an adaptive quadrature method for Calabi--Yau geometry.
In the toric/CICY formulation, one chooses K\"ahler cone generators $J^{(\alpha)}$, constructs bases of sections $s^{(\alpha)}_j$, maps the ambient toric coordinates into projective spaces, and equips these spaces with auxiliary Fubini--Study potentials
\begin{equation}
K^{(\alpha)}=\log\!\left(\sum_{i,j}\bar s^{(\alpha)}_i\,H^{ij}\,s^{(\alpha)}_j\right) ~.
\end{equation}
The na\"{\i}ve choice $H=\mathbf 1$ gives a known measure, but it can place too many points in some regions and too few in others.
Following the Keller--Lukic idea, one instead varies the auxiliary Hermitian data, or equivalently uses several sampling regions, so that new points are pushed into regions, typically with high curvature, that were undersampled by the first pass~\cite{Keller:2009}.
Near a conifold point of the quintic, the dominant error in the Euler number computation is numerical integration rather than the neural metric itself; improved point sampling reduces the error from roughly $15\%$ to roughly $2\%$.
As well, there is a multiple order of magnitude improvement in the metric approximation and a $60\%$ improvement in the harmonic $(2,1)$-form approximation when the point cloud is improved~\cite{Ruehle:Pollica2025,Lust:2026mys}.
This is precisely the regime in which better point selection should pay off: Ricci-flat metrics, harmonic forms, curvature invariants, and warped sources can have localized structure, and an adaptive sampler can direct training points toward the regions where the geometric residuals actually live~\cite{Anderson:2020hux,Larfors:2021pbb,Larfors:2022nep}.

\section{Heterotic compactification and the low-energy effective action}
\label{s:hetS}
\subsection{The central observables}

Consider the $E_8\times E_8$ heterotic string compactified on a smooth \CY threefold $X$ with holomorphic vector bundle $V$~\cite{Candelas:1985en}.
Chiral matter fields arise from bundle-valued cohomology groups such as $H^1(X,V_R)$, and their internal zero modes are represented by $(0,1)$-forms $\nu_I$ that are $\bar\partial_A$-harmonic with respect to the Ricci-flat metric $g$ on $X$ and the Hermitian Yang--Mills (HYM) connection $A$ on $V$.
The holomorphic cubic couplings are schematically
\begin{equation}
\lambda_{IJK} = \int_X \Omega \wedge \Tr\big(\nu_I\wedge \nu_J\wedge \nu_K\big) ~,
\label{eq:holomorphic-yukawa}
\end{equation}
while the matter K\"ahler metric is determined by the overlap matrix
\begin{equation}
Z_{I\bar J} = \frac{1}{\cV} \int_X \dd\mathrm{Vol}_{g}\; \langle \nu_I,\nu_J\rangle_{g,h_V}~.
\label{eq:matter-metric}
\end{equation}
Here $\langle\cdot,\cdot\rangle_{g,h_V}$ denotes the pointwise Hermitian inner product on bundle-valued forms induced by the Ricci-flat metric $g$ on $X$ and the Hermitian bundle metric $h_V$; the displayed formula suppresses convention dependent trace and normalization factors.
The physical Yukawa couplings are not the holomorphic quantities in~\eqref{eq:holomorphic-yukawa} themselves, but the canonically normalized couplings obtained after diagonalizing and rescaling the kinetic terms:
\begin{equation}
Y^\text{phys}_{IJK} = e^{K/2}\, \lambda_{ABC}\,(Z^{-1/2})^{A}{}_{I}(Z^{-1/2})^{B}{}_{J}(Z^{-1/2})^{C}{}_{K} ~.
\label{eq:physical-yukawa}
\end{equation}
This distinction is crucial.
For heterotic compactifications one ultimately needs the Ricci-flat Calabi--Yau metric, the HYM connection on the bundle, and normalized harmonic representatives of the relevant bundle-valued cohomology classes; holomorphic Yukawa couplings alone are only part of the story.
In favorable cases the holomorphic couplings can be obtained quasi-topologically, but the matter metrics and hence the physical masses and mixings depend on genuinely non-holomorphic differential geometry data.
The Ricci-flat metric enters essentially in two respects: in the Hodge star operator and in knowing the harmonic forms.

This program has now been carried out explicitly in a number of standard embedding models, for which the holomorphic vector bundle solving the HYM equations is the tangent bundle~\cite{Candelas:1985en,Strominger:1985it}.
In this context, parts of the matter normalization problem are tied to the Weil--Petersson metric on complex structure moduli space, which can be computed using period integrals or the Kodaira--Spencer map without first constructing the full Ricci-flat metric~\cite{Candelas:1990pi,Candelas:1990rm}.
This led to direct computations of normalized Yukawa couplings in explicit examples~\cite{Butbaia:2024tje} and then to higher precision calculations across several models, with natural hierarchies emerging after canonical normalization~\cite{Berglund:2024uqv}.
The numerical values for the Yukawa couplings match when these are calculated using machine learned harmonic forms constructed from the machine learned numerical Ricci-flat metric.
This supports using the same approximations in non-standard embeddings, where the special geometry shortcuts are generally not available.

A first direct numerical computation of physical Yukawa couplings in a genuinely non-standard heterotic compactification was achieved for line bundle models with Abelian internal flux~\cite{Constantin:2024yxh}.
Complementary bottom-up and top-down studies have clarified the flavor story in split bundle and line bundle settings.
For example,~\cite{Constantin:2024yaz} analyzes string inspired Froggatt--Nielsen structures motivated by heterotic split bundles, while~\cite{Constantin:2025vyt} exhibits explicit heterotic line bundle models that reproduce the observed quark and charged lepton masses and quark mixing, subject to reasonable assumptions about the moduli.
The differential geometric program has also been extended to a non-Abelian bundle example by a method that is general with respect to the rank and structure group of the holomorphic bundle, provided one can explicitly construct the relevant section bases~\cite{Mishra:2025xkr}.
In parallel, approximate analytic Ricci-flat metrics with explicit K\"ahler moduli dependence have begun to appear~\cite{Constantin:2026bky}, which is an important step toward making these calculations genuinely moduli dependent.

This is still not the most general heterotic case.
Current results cover the standard embedding and selected non-standard settings, most notably line bundle constructions, split bundle inspired effective theories, and now particular non-Abelian examples on smooth K\"ahler Calabi--Yau compactifications.
They do not yet amount to a fully general treatment of arbitrary heterotic compactifications, nor do they yet provide a uniform handle on the simultaneous dependence on complex structure, K\"ahler, and bundle moduli in complete generality.

The same logic applies to metrics on the moduli space of Calabi--Yau geometries.
For example, the Weil--Petersson metric on complex structure moduli space is obtained from
\begin{equation}
K_{\rm cs}=-\log\!\left(i\int_X \Omega\wedge\bar\Omega\right),
\qquad
G^{\WP}_{\alpha\bar\beta}=\partial_\alpha\partial_{\bar\beta}K_{\rm cs}.
\end{equation}
Equivalently, with standard orientation and normalization conventions, it may be written schematically as
\begin{equation}
G^{\WP}_{\alpha\bar\beta} \sim -\frac{\int_X \chi_\alpha \wedge \bar\chi_{\bar\beta}}{i\int_X \Omega \wedge \bar\Omega},
\end{equation}
where $\chi_\alpha\in H^{2,1}(X)$ are harmonic representatives.
K\"ahler moduli metrics are similarly given by overlaps of harmonic $(1,1)$-forms, up to the usual convention dependent normalizations.
In the standard embedding, these metrics are directly tied to matter normalizations, which is why numerical calculations of physical Yukawa couplings were possible there first~\cite{Butbaia:2024tje,Berglund:2024uqv}.
More generally, however, making the moduli dependence explicit remains hard: even when accurate numerical metrics are available at fixed moduli, one still lacks a completely general analytic description of their variation across moduli space.
Recent work on analytic approximations with explicit K\"ahler moduli dependence is an important step, but it is again only a step toward the fully general problem~\cite{Constantin:2026bky}.

\subsection{What can now be computed}
Once numerical approximations to $\num{g}$, $\num{A}$, and the relevant harmonic representatives are known, one can begin to compute quantities that were previously inaccessible at the level of the canonically normalized four-dimensional theory.
What is feasible in practice depends strongly on the compactification class: the standard embedding is currently the most mature case, selected non-standard embeddings are now accessible, while a completely general treatment remains out of reach.
Within that caveat, one can attack:
\begin{itemize}
  \item \textbf{Normalized Yukawa couplings and flavor textures:} One can study not just whether a coupling is present, but whether hierarchical structures arise already at the holomorphic level or only after canonical normalization, and whether approximate flavor symmetries survive the passage to the physical basis.
  \item \textbf{Masses:} These include Kaluza--Klein masses from Laplace and Dirac spectra, vector-like masses generated by bundle moduli or Wilson lines, and ultimately visible sector fermion masses once moduli stabilization and Higgsing are specified.
  \item \textbf{Higher-dimension operators:} Quartic superpotential terms, proton decay operators, neutrino operators, and other effective couplings depend on products of normalized internal wavefunctions and, beyond leading order, on Green's operators on $X$.
  \item \textbf{Threshold corrections:} One-loop thresholds and gauge kinetic corrections depend on spectra and determinants of elliptic operators, so numerical spectral geometry becomes directly relevant.
  \item \textbf{Wavefunction localization and kinetic mixing:} Numerical wavefunctions make it possible to test directly whether flavor hierarchies are associated with localized or quasi-localized internal profiles, or instead are dominated by kinetic mixing after canonical normalization.
\end{itemize}

\subsection{Important open problems}

The most important open phenomenological problems that now become concrete are:
\begin{enumerate}[label=(\roman*)]
  \item \textbf{Full flavor physics in explicit heterotic Standard Models:}
  The immediate frontier is no longer just computing a few Yukawa couplings, but entire Yukawa matrices, kinetic mixings, higher order corrections, and eventually CKM/PMNS data in compactifications with realistic spectra.
  Examples include heterotic GUT and MSSM-like constructions such as~\cite{Braun:2005nv,Bouchard:2005ag}, where the algebraic and topological data are well developed but physical normalized couplings still require metric input.
  \item \textbf{Large scans of physical couplings:}
  We already know how to scan topological models; the next step is to scan metric normalized observables over large model sets and ask which flavor patterns are generic, rare, or impossible.
  \item \textbf{Selection rules versus normalization effects:}
  Many heterotic Yukawa textures are understood algebraically, but the competition between texture zeros, kinetic mixing, and wavefunction localization is still poorly mapped.
  Precision numerics can separate topological zeros from merely tiny normalized couplings.
  \item \textbf{Higher order operators and precision EFT:}
  The field has mostly focused on holomorphic cubic couplings in the $E_6$ language, including $\mathbf{27}^3$ and $\overline{\mathbf{27}}^3$ sectors. Their holomorphic pieces can be protected or quasi-topological in special embeddings, but the physical normalized couplings, and the mixed $\mathbf{27}{\cdot}\overline{\mathbf{27}}{\cdot}\mathbf{1}$ and $\mathbf{1}^3$ couplings needed for the lowest-order $E_6$ effective action, require matter metrics~\cite{Hubsch:2024cym}.
  To ultimately make contact with the Standard Model, constructive string phenomenology also needs quartic operators, dimension five proton decay operators, neutrino mass operators, and moduli dependent corrections with error bars.
  \item \textbf{Toric Calabi--Yau geometries:} While various codes, \textit{e.g.}, \texttt{cymetric}~\cite{Larfors:2021pbb}, obtain approximations to the Ricci-flat metric on toric Calabi--Yau spaces this, the performance is notably worse in comparison to the complete intersection geometries.
  There are only $7890$ complete intersection Calabi--Yau threefolds.
  Starting from a triangulation of a four-dimensional reflexive polytope, we can construct a toric variety in which the anticanonical hypersurface is a (possibly) singular Calabi–Yau variety~\cite{Batyrev:1993oya}.
  There are $473,800,776$ reflexive polytopes to start from~\cite{Kreuzer:2000xy} and an unknown number of triangulations of these; see, however~\cite{MacFadden:2024him,Berglund:2024reu,MacFadden:2025ssx}.
  Extending the framework to F-theory, we can supplement the known five-dimensional reflexive polytopes~\cite{Scholler:2018apc} with machine learned examples~\cite{Berglund:2023ztk} and thereby construct new fourfolds for which similar considerations arise.

\end{enumerate}

\section{Moduli stabilization, supersymmetry breaking, and dark sectors}
\label{s:stab}
This section collects the compactification problems in which the Calabi--Yau metric enters most directly through the four-dimensional effective potential and its scalar spectrum: $\alpha'$ corrections and quantum Calabi--Yau metrics, supersymmetry breaking and soft terms, de~Sitter model building, moduli stabilization, and axion and dark-sector physics.

\subsection{\texorpdfstring{$\alpha'$}{alpha'} corrections and quantum Calabi--Yau metrics}
\label{s:alpha-corrections}
The numerical Ricci-flat \CY metric is the leading background in the two-derivative, large volume expansion, not the full metric of the finite volume string vacuum.
An important early step in this direction was the worldsheet analysis of Nemeschansky and Sen, who showed that the appearance of higher-loop corrections does not destroy conformal invariance for strings on a Calabi--Yau target: even though the corrected background metric is no longer exactly Ricci-flat, one can still maintain a conformally invariant supersymmetric sigma model, with the corrected metric related to the Ricci-flat one by a (generally non-local) field redefinition~\cite{Nemeschansky:1986yx}.
This picture was later reformulated in target space: higher derivative corrections deform an $SU(3)$-holonomy Calabi--Yau background to a supersymmetric $SU(3)$-structure background for which the physical metric need not remain Ricci-flat, while $c_1(X)=0$ is preserved; moreover, in the $B_{ab}=0$ sector, an explicit first order Ans\"atz $J'=J+da$ in which the scalar controlling the deformation satisfies a Poisson equation gives a perturbative route away from the leading Ricci-flat metric~\cite{Becker:2015wga}.
Read in the present large volume setting, these results support treating the numerically determined Ricci-flat metric as the zeroth-order term and then solving iteratively for its $1/\mathcal{V}$-suppressed quantum corrections.

The Hull--Strominger system is the prototypical heterotic example of this programme.
Anomaly cancellation and supersymmetry require the Bianchi identity
\begin{equation}
\dd H = \frac{\alpha'}{4}\big(\Tr R\wedge R - \Tr F\wedge F\big) ~,
\label{eq:bianchi}
\end{equation}
and more generally a torsional geometry in which the internal space carries a complex structure and a holomorphic bundle, but the Hermitian metric need not be K\"ahler and the $H$-flux sources the torsion~\cite{Hull:1985ab, Hull:1986kz, Strominger:1986uh, Li:1986hh, Dasgupta:1999ss, Goldstein:2002pg, Fu:2006vj, Becker:2005nb, Melnikov:2014ywa, Candelas:2016usb}.
This is the clearest heterotic illustration of why the numerical Ricci-flat metric is only a starting point: the full string background requires solving simultaneously for the torsion, the $H$-flux, and the corrected bundle data.
Even if the exact background is not K\"ahler, the large volume Ricci-flat metric is the natural seed from which to begin a controlled iterative, flow-based, or Newton-type numerical solution of the full torsional system.

The currently available Ricci-flat metrics apply to the $\mathcal{V}/(2\pi)^6\alpha^{\prime\, 3}\to\infty$ limit of the compactification.
In isotropic large volume scaling it is useful to factor out the overall size of the threefold and write schematically
\begin{equation}
 g_{m\bar n} = \mathcal{V}^{1/3}\,\hat g_{m\bar n} ~, \qquad
 \hat g_{m\bar n} = \hat g^{(0)}_{m\bar n} + \mathcal{V}^{-1}\hat g^{(1)}_{m\bar n} + \mathcal{V}^{-2}\hat g^{(2)}_{m\bar n} + \ldots ~,
\end{equation}
with analogous expansions for the gauge bundle and the remaining background fields.
The precise powers can depend on the sector, on anisotropic scalings, and on the string theory under discussion; in type II examples the first correction to the effective action is often discussed at order $\alpha'^3$, while heterotic backgrounds receive order-$\alpha'$ corrections tied to the Bianchi identity~\eqref{eq:bianchi}.
The key point is universal: the Ricci-flat metric is only the zeroth-order term.
This perspective is made particularly explicit in~\cite{Fraser-Taliente:2024etl}, where numerical Ricci-flat metrics are used both to sharpen the usual large volume criterion by pointwise curvature tests and to compute the first correction to the metric itself.

Once a numerical seed is known, the next step is not merely to evaluate observables on $g^{(0)}$, but to solve perturbatively for the corrected fields,
\begin{equation}
 g = g^{(0)} + \epsilon g^{(1)} + \epsilon^2 g^{(2)} + \cdots ~, \qquad
 A = A^{(0)} + \epsilon A^{(1)} + \cdots ~.
\end{equation}
Here $\epsilon$ denotes the relevant small parameter in the chosen large volume or weak coupling regime. After stripping off the overall size, this becomes an inverse volume expansion of the corrected background.
In type IIB compactifications,~\cite{Bonetti:2016dqh} showed that the leading $\alpha'^3$ correction already modifies the internal background metric, introduces a non-trivial Weyl factor, and deforms the internal space away from a strict Ricci-flat Calabi--Yau metric toward an almost-\CY $SU(3)$-structure background.
For present purposes, this is the cleanest prototype of a ``quantum Calabi--Yau metric'': the corrected target space metric data obtained order by order from the Ricci-flat seed, even when the corrected geometry is no longer exactly Ricci-flat.

The numerical route to quantum \CY metrics in a $1/\mathcal{V}$-expansion is conceptually straightforward.
One first computes $g^{(0)}$ and, when relevant, $A^{(0)}$.
One then linearizes the corrected field equations about this background.
At each order one obtains elliptic equations for the corrections $g^{(n)}$, $A^{(n)}$, $B^{(n)}$, and $\phi^{(n)}$, with source terms built from lower-order data: local curvature polynomials, $\Tr R\wedge R$, $\Tr F\wedge F$, flux bilinears, and localized source contributions.
In the type IIB $\alpha'^3$ example analyzed in~\cite{Fraser-Taliente:2024etl}, the metric correction can be written in terms of a corrected K\"ahler potential satisfying a Poisson equation whose source is essentially the six-dimensional Euler density with its zero mode removed.
The same strategy should extend, in favorable heterotic settings and after gauge fixing, to perturbation theory around suitable seeds: once the zeroth-order metric and bundle are known numerically, the Bianchi identity and supersymmetry conditions determine explicit source terms order by order, and the corrected geometry can be accessed by solving the resulting sourced PDEs on the compact manifold.

Several quantities that are hard to access analytically become straightforward once the numerically computed $g$ and $A$ are available: pointwise norms of $R$, $F$, and $H$; local curvature hotspots where the derivative expansion is least trustworthy; explicit representatives for the source term in~\eqref{eq:bianchi}; corrected moduli space metrics; corrected matter K\"ahler metrics; and corrected kinetic operators whose spectra determine KK masses and wavefunctions.
In type IIB compactifications, the same technology can be used to compute leading $\alpha'^3$ or loop corrected source terms more faithfully than the usual volume scaling estimates~\cite{Becker:2002nn,Bonetti:2016dqh,Fraser-Taliente:2024etl}.

\subsubsection{Open problems}

Relevant open problems include:
\begin{enumerate}[label=(\roman*)]
  \item \textbf{Systematic $1/\mathcal{V}$ algorithms for quantum \CY metrics:} One wants a practical scheme that takes a numerical Ricci-flat seed and returns the successive corrections $g^{(1)}$, $g^{(2)}$, \ldots, together with the corresponding corrections to the bundle, $B$-field, and dilaton, with gauge fixing and zero-mode subtraction handled cleanly.
  \item \textbf{Corrected matter metrics and physical Yukawa couplings:} It is already known that heterotic moduli metrics receive $\alpha'$ corrections~\cite{Candelas:2016usb}.
  The obvious next question is how much these corrections change physical Yukawa couplings, soft terms, and flavor hierarchies in realistic vacua.
  At Gepner or other intrinsically stringy points, classical large radius geometry is not by itself the right object. One must specify whether the calculation is being performed with periods, the exact conformal field theory/Zamolodchikov metric, or a large radius continuation; this distinction is especially important when discussing physical normalizations in the $\overline{\mathbf{27}}^3$ sector.
  \item \textbf{Local control of the derivative expansion:} Instead of relying on global volume arguments alone, one should compute local dimensionless parameters such as $\alpha'|R|$ and identify where compactifications are close to breaking down.
  \item \textbf{Continuation from \CY to corrected $SU(3)$-structure geometry:} The machine learning study of $SU(3)$-structure metrics was initiated in~\cite{Anderson:2020hux}.
  The type IIB $\alpha'^3$ correction already shows that the corrected background need not remain a strict Ricci-flat \CY metric~\cite{Bonetti:2016dqh}.
  Numerical Ricci-flat metrics should therefore be used as initial data for anomaly flow, balanced flow, or Newton-type methods that continue toward the corrected supersymmetric background.
  \item \textbf{Explicit compact numerical solutions of the Hull--Strominger system:} The field has many structural results, but very few genuinely explicit compact solutions with all fields under simultaneous control --- metric, bundle, $B$-field, dilaton, and $H$-flux all satisfying~\eqref{eq:bianchi} and the remaining supersymmetry conditions.
\end{enumerate}

\subsection{Supersymmetry breaking and soft terms}
\label{s:SuSyB}
Phenomenology ultimately depends not only on the supersymmetric effective action but on how supersymmetry is broken.
In four-dimensional $\mathcal{N}=1$ supergravity the soft terms are controlled by derivatives of the K\"ahler potential, the matter metrics and the gauge kinetic functions; standard formul\ae\ are reviewed~\cite{Brignole:1997dp}.
Schematically,
\begin{align}
(m^2)_{I\bar J} &= (m_{3/2}^2+V_0) Z_{I\bar J} - F^A\bar F^{\bar B} R_{A\bar B I\bar J} ~, \\
A_{IJK} &= F^A\partial_A \log\!\left(\frac{e^K\lambda_{IJK}}{Z_I Z_J Z_K}\right), \\
M_a &= \frac{1}{2\,\Re f_a}F^A\partial_A f_a ~.
\end{align}
The $A$-term formula is written in the common diagonal matter metric notation.
In a general basis one must use the full matrix valued matter metric, its Chern connection on the matter bundle over moduli space, and the corresponding curvature tensor.
Thus the internal geometry enters twice: first through the normalized matter metrics $Z_{I\bar J}$, and second through the moduli space geometry that determines the curvature tensor $R_{A\bar B I\bar J}$.
In particular, even when the holomorphic Yukawa couplings are fixed by cohomological data, the physical Yukawa couplings, A-terms and scalar masses still depend on wavefunction normalization and hence on the Calabi--Yau metric, bundle metric and harmonic representatives~\cite{Blesneag:2018mfm,Butbaia:2024tje,Constantin:2024yxh}.

This immediately opens several concrete research directions.
One can ask whether hidden sector supersymmetry breaking leads to flavor universal soft masses or to dangerous flavor non-universality; whether approximate sequestering occurs in explicit compact models or is spoiled by the full compact geometry; whether kinetic mixing aligns or misaligns A-terms with Yukawa couplings; and whether geometrically small wavefunction overlaps suppress unwanted couplings.
From the effective field theory viewpoint these questions are controlled by the curvature of the scalar manifold, and in generic Calabi--Yau compactifications this curvature is not fixed by symmetry alone.
The analysis of Farquet and Scrucca shows that soft scalar masses depend sensitively on departures from special coset geometries and that only a mild form of sequestering can be expected without additional structure~\cite{Farquet:2012sgm}.

The heterotic setting is especially interesting because sequestering is subtle: there are no separated visible and hidden branes in the same simple sense as in some type II models, so the answer depends on global geometry and bundle structure rather than on a cartoon of spatial separation.
More generally, even in settings where sectors appear geometrically isolated, moduli mediated and nonperturbative effects can reintroduce visible/hidden couplings; geometric distance by itself is not a theorem of sequestering~\cite{Berg:2010ha}.
In heterotic compactifications, one therefore wants basis independent calculations of the actual matter metrics, field space curvatures and overlap integrals, rather than qualitative claims that one sector is ``far away.''

There is some analytic control in special regimes.
For heterotic compactifications with Abelian internal gauge flux, the matter field K\"ahler metric can localize in regions of the internal space, making approximate calculations possible even without complete knowledge of the exact Ricci-flat background~\cite{Blesneag:2018mfm}.
At the same time, the moduli kinetic terms themselves receive stringy corrections: in heterotic Calabi--Yau compactifications the leading corrections to the four-dimensional K\"ahler potential can already appear at order $(\alpha')^2$ in generic situations, thereby modifying the no-scale structure and hence the soft terms induced after supersymmetry breaking~\cite{Anguelova:2010ed}.
For phenomenology, this means that both the tree-level internal geometry and its quantum corrections can feed into the final soft spectrum.

Numerical geometry now makes this program more concrete.
Recent work has computed normalized Yukawa couplings and even quark masses in explicit heterotic Calabi--Yau compactifications by numerically approximating the Ricci-flat metric, Hermitian Yang--Mills bundle metrics and harmonic forms~\cite{Butbaia:2024tje,Constantin:2024yxh}.
The same toolkit should make it possible to go beyond qualitative discussions of flavor, sequestering and alignment and toward explicit predictions for soft scalar masses, A-terms and gaugino masses in compact examples.
The open problem here is to turn the geometric input data of a compactification into a numerically controlled, basis independent computation of the full soft term sector.

\subsubsection{Open problems}
We thus have the following questions to consider:
\begin{enumerate}[label=(\roman*)]
\item \textbf{Basis independent soft terms in explicit compactifications:}
Given a compact heterotic vacuum with numerical approximations to the Ricci-flat metric, the Hermitian Yang--Mills connection, and the relevant harmonic bundle-valued forms, compute the full basis independent soft-term data $(m^2)_{I\bar J}$, $A_{IJK}$, and $M_a$, with numerical error estimates propagated from the internal geometry to the four-dimensional EFT.

\item \textbf{Flavor universality versus geometric flavor violation:}
Determine in explicit compact models whether hiddensector supersymmetry breaking produces flavor-universal scalar masses, or whether kinetic mixing, wavefunction overlap, and scalar manifold curvature lead to dangerous flavor violation after canonical normalization.

\item \textbf{Quantitative tests of sequestering:}
Test sequestering in compact geometries by computing visible-hidden overlap integrals, matter-metric curvatures, and moduli mediated couplings, rather than relying only on geometric separation or local intuition.

\item \textbf{Moduli dependence of matter metrics:}
Compute the dependence of matter K\"ahler metrics on complex structure, K\"ahler, and bundle moduli, and from it extract the mixed curvature tensor $R_{A\bar B I\bar J}$ that enters the scalar soft masses.

\item \textbf{Alignment of A-terms and physical Yukawa couplings:}
Determine whether A-terms align with physical Yukawa matrices in explicit vacua, and identify the geometric conditions under which approximate alignment, suppression of CP phases, or suppression of flavor violation occurs.

\item \textbf{$\alpha'$ corrections to the soft spectrum:}
Extend soft-term calculations beyond the leading Ricci-flat background by incorporating the $\alpha'$ corrections to the K\"ahler potential, matter metrics, Hermitian Yang--Mills data, and gauge kinetic functions that can shift scalar masses, A-terms, and gaugino masses.
\end{enumerate}

\subsection{de~Sitter model building}
\label{s:deSitter}
We live in an asymptotically de~Sitter spacetime.
This is notoriously difficult to realize in quantum gravity.
Cosmological model building of the present epoch in string theory is an example of a problem where percent level geometric effects can decide the answer.
Proposed constructions in type II superstring theory balance flux superpotentials, non-perturbative terms, $\alpha'$ corrections, warping, and uplift effects.
All of these are exponentially or parametrically sensitive to cycle volumes, canonical normalization, and corrections to the effective potential.
In other words, they are sensitive to the Calabi--Yau metric.

In type IIB flux compactifications the Ricci-flat Calabi--Yau metric is not only a zeroth-order approximation to the internal geometry; it is the geometric data with respect to which the warped solution is constructed.
Explicitly, in the Giddings--Kachru--Polchinski setup~\cite{Giddings:2001yu}, the ten-dimensional Einstein frame metric is written as
\begin{equation}
ds_{10}^{2} = e^{2A(y)} \eta_{\mu\nu} dx^\mu dx^\nu + e^{-2A(y)} \widetilde g_{m\bar n}(y)\, dy^m d\bar y^{\bar n} ~,
\qquad \text{Ric}(\widetilde g)=0 ~,
\label{eq:gkpmetric}
\end{equation}
where $\widetilde g$ is the unwarped Ricci-flat metric on the Calabi--Yau threefold $X$.
The warp factor $A(y)$ encodes the backreaction of three-form fluxes, localized D$3$-brane charge, and orientifold sources.
The Ricci-flat metric $\widetilde g$ supplies the Laplacian, Hodge star, volume form, Green's function, and pointwise norms that enter the warped equations.

The flux background is specified by
\begin{equation}
G_3 = F_3-\tau H_3 ~,
\qquad \widetilde \star_6 G_3 = i G_3 ~,
\label{eq:isd}
\end{equation}
together with the self-dual five-form
\begin{equation}
\widetilde F_5 = (1+\star_{10})\, d\alpha \wedge dx^0\wedge dx^1\wedge dx^2\wedge dx^3 ~,
\end{equation}
with $\alpha$ a function of the compact space.
The complex structure moduli and the axio-dilaton are fixed by the F-flatness equations following from the Gukov--Vafa--Witten superpotential~\cite{Gukov:1999ya}
\begin{equation}
W = \int_X \Omega \wedge G_3 ~,
\qquad D_\tau W=0 ~,
\qquad D_i W = \int_X D_i\Omega \wedge G_3 =0 ~.
\label{eq:fluxfterms}
\end{equation}
Cohomologically, these equations can be expressed in terms of periods of $\Omega$, but a local warped solution requires more than the periods.
One must know harmonic representatives of the $(2,1)$-forms with respect to $\widetilde g$, because the local source for the warp factor is the pointwise flux density.
In the conventions adapted to the GKP solution, the warp factor satisfies a Poisson equation in which the Laplacian depends on the Ricci-flat metric.

This gives a natural computational hierarchy.
First, one solves for the Ricci-flat Calabi--Yau metric $\widetilde g$ in the desired K\"ahler class.
One then uses periods to determine the flux vacuum and constructs metric dependent harmonic representatives of the relevant $(2,1)$-forms.
Finally, one evaluates the source terms and solves for the warp factor~\cite{Giddings:2001yu}:
\begin{equation}
\widetilde \nabla^2 e^{4A} = i e^{2A} \frac{G_{mnp} (\widetilde \star_6 \overline{G}^{mnp})}{12\;\text{Im}\,\tau}+2e^{-6A}(\partial_m e^{4A})(\partial^m e^{4A})+2\kappa_{10}^2e^{2A}T_3\rho_3^\text{loc} ~,
\end{equation}
where $\rho_3^\text{loc}$ denotes the effects of localized D$3$-branes.
The resulting warped geometry determines physical quantities such as the four-dimensional Planck mass, warped Kaluza--Klein scales, normalized matter wavefunctions, local Yukawa couplings, and the redshift of energy scales in strongly warped regions.
Near conifold points or localized sources, where the geometry can be nearly singular, accurate approximations to $\widetilde g$ and careful point sampling become especially important.
The preliminary efforts at finding solutions that were reported in~\cite{Ruehle:Pollica2025} are extended in~\cite{Lust:2026mys}.

For candidate KKLT-type vacua, one moreover wants accurate divisor volumes, warped throat data, moduli space metrics, and a trustworthy Hessian after uplift~\cite{Kachru:2003aw,Giddings:2001yu,McAllister:2024lnt}.
A Klebanov--Strassler throat is not just a convenient local toy model in this story.
It is often the sector that generates a parametrically small redshift, sequesters supersymmetry breaking from the bulk, and makes an anti-D3 uplift small enough to compete with non-perturbative stabilization rather than simply overpower it~\cite{Giddings:2001yu,Klebanov:2000hb,Kachru:2003aw}.
In the recent explicit candidate vacua of~\cite{McAllister:2024lnt}, each example contains a Klebanov--Strassler throat with a single anti-D3-brane, so the throat is part of the construction itself rather than an auxiliary afterthought.
The local dynamics of anti-branes in such a throat is also tied to the metastability analysis of Kachru, Pearson, and Verlinde~\cite{Kachru:2002gs}.
More broadly, Klebanov--Strassler throats provide explicit local sectors with tunable hierarchies and strongly warped tips, and statistical arguments suggest that throat regions are not rare in large ensembles of flux vacua~\cite{Hebecker:2006jc}.

For large volume scenarios (LVS), one instead wants reliable control of the competition between the leading $\alpha'$ correction, non-perturbative terms, loop effects, and uplift sectors~\cite{Balasubramanian:2005zx}.
But here too warped local sectors can matter, because uplift energies, soft masses, and canonical normalization can depend sensitively on how local sources are embedded into the global metric.
In either KKLT or LVS, the question is not only whether a minimum exists in naive moduli coordinates, but whether it survives after passing to canonically normalized fields, after throat/bulk mixing is included, and after the dominant computable geometric corrections are taken into account.

This is where numerical \CY geometry could add genuinely new input.
Following the curvature diagnostics emphasized in~\cite{Berglund:2022gvm}, one can use scalar invariants such as the Kretschmann scalar
\begin{equation}
K \equiv \big\|\text{Riem}\big\|^2_{\widetilde{g}}
  = R_{m\bar np\bar q}R^{m\bar np\bar q}
\end{equation}
to identify where the internal metric develops large local curvature.
In practice, one can sample $K$ over the manifold, examine the high curvature tail of its distribution, and then use clustering or persistent homology methods to isolate localized regions of unusually large curvature.
In the examples studied in~\cite{Berglund:2022gvm}, high curvature regions cluster in the vicinity of (nearly) singular points.
For the model building applications at issue here, that observation is a useful diagnostic rather than an identification theorem. In a compact geometry with conifold degenerations, localized high curvature regions are natural candidates for neighborhoods that glue onto warped throats or contain shrinking cycles, but period data, explicit cycle representatives, warping, and flux quantization are needed before a throat or uplift sector has been identified.

That, in turn, suggests a possible first principles strategy for flux compactifications.
Rather than specifying a throat only abstractly in terms of periods and topological data and then attaching a local uplift sector by hand, one would like to combine period calculations, cycle localization, warping estimates, and curvature diagnostics in the numerical metric.
The fluxes are supported on cycles, not at points, and the tip at which an anti-brane would sit is determined by the warped geometry and the full backreaction problem.
High-curvature loci isolated by $K$ may therefore be good places to look, but not by themselves sufficient evidence for a controlled uplift.
This does not mean that large curvature by itself validates the ten-dimensional supergravity description; if anything, it flags the regions where one must be most careful.
Using knowledge of the metric, we can also consider perturbations to the Klebanov--Strassler throats~\cite{Baumann:2010sx,Gandhi:2011id}.
Because it provides a quantitative localization criterion, numerical metrics could help turn the slogan ``put fluxes on the cycles that make a throat, and put the anti-brane at the tip'' into a globally defined diagnostic rather than a partly local Ansatz.

Numerical \CY geometry sharpens the EFT data entering de~Sitter constructions, and recent work has made explicit leading order candidate vacua in concrete constructions~\cite{McAllister:2024lnt}.
The existence of fully controlled metastable de~Sitter vacua remains unsettled, because unknown subleading corrections, source backreaction, and genuinely stringy issues are still central~\cite{Sethi:2017phn, Dine:2020vmr, Danielsson:2018ztv, Bena:2011wh}.
Numerical metrics do not solve these issues by themselves.
What they do is something equally important: they let one distinguish which ambiguities are truly string-theoretic and which were merely consequences of crude geometric approximations, and they offer a way to replace vague local statements about ``a throat somewhere in the compactification'' by a more explicit global picture of where the relevant localized sectors actually live.

\subsubsection{Open problems}
Open problems relevant to de~Sitter space and the cosmological constant include:
\begin{enumerate}[label=(\roman*)]
\item \textbf{Metric corrected tests of candidate de~Sitter vacua:}
Determine which candidate explicit de~Sitter vacua survive after all currently computable metric effects are included, especially throat/bulk gluing effects and canonical normalization in the warped geometry.

\item \textbf{Numerical diagnostics for compact throats:}
Develop practical numerical diagnostics for locating Klebanov--Strassler-like regions in compact models, combining period data, warping estimates, and curvature invariants such as the Kretschmann scalar.

\item \textbf{High-curvature regions and flux-supported cycles:}
Test whether the high curvature regions isolated by $K$ correlate with shrinking cycles that support flux and with the preferred locations of localized uplift sectors.

\item \textbf{Backreaction and antibrane polarization in compact models:}
Quantify source backreaction, flux induced warping, and antibrane polarization in explicit compact models rather than in local throat approximations alone.

\item \textbf{Heterotic de~Sitter constraints with numerical geometry:}
Understand whether heterotic de~Sitter model building can be sharpened in an analogous way, especially in settings with $D$-terms, threshold corrections, and $\alpha'$-induced potentials.
\end{enumerate}

\subsection{Moduli stabilization}
\label{s:stabM}
A numerical Ricci-flat metric is equally valuable for vacuum selection. In heterotic compactifications on smooth \CY backgrounds, moduli stabilization is notoriously difficult because there is no generic analogue of the type IIB flux superpotential that fixes all geometric moduli.
One instead combines several mechanisms: bundle holomorphy constraints, D-terms from anomalous $U(1)$ factors, worldsheet instantons, gaugino condensation, threshold corrections, and sometimes a move away from the strictly K\"ahler regime toward torsional backgrounds~\cite{Anderson:2010mh,Anderson:2011cza,Strominger:1986uh,McAllister:2023vgy}.
The point of a numerical metric is not merely to improve a picture of the compact space.
It turns the stabilization problem from an algebraic existence question into a quantitative question about the four-dimensional effective theory.
The present generation of numerical Calabi--Yau methods, beginning with balanced metric and energy functional approaches and now including moduli dependent machine learning constructions, makes this program concrete for hypersurfaces, quotients, and complete intersections.

The reason the numerical metric matters is that stabilization is not just about writing a formal potential; it is about computing the actual scalar potential in canonically normalized fields.
Several indispensable ingredients are metric dependent:
\begin{align}
\cV &= \frac16\int_X J\wedge J\wedge J \,, &
\Vol(D) &= \frac12 \int_D J\wedge J \,, &
\Vol(C) &= \int_C J\,.
\end{align}
For a numerical metric, the Monge--Amp\`ere residual is more properly measured by a normalized comparison between the metric volume form $\omega_{\num{g}}^3/3!$ and the holomorphic volume form $i\,\Omega\wedge\bar\Omega$ after fixing their total volumes. This residual is a diagnostic for the numerical approximation, not a separate metric independent expression for the physical volume.
At leading order, volumes of calibrated holomorphic cycles are often expressible in terms of intersection numbers and K\"ahler parameters.
However, a Ricci-flat metric supplies the harmonic representatives, Hodge star, local volume form, and curvature distribution needed to convert this cohomological data into physical kinetic terms, threshold corrections, and error estimates.
In particular, the moduli space metric can be written schematically as
\begin{equation}
G^{(K)}_{\alpha\bar\beta}\sim \frac{1}{\cV}\int_X \omega_\alpha\wedge *\omega_\beta \,,
\qquad
G^{(\text{cs})}_{i\bar j}\sim
\frac{\int_X \chi_i\wedge *\overline{\chi}_{\bar j}}{i\int_X \Omega\wedge \overline\Omega} \,,
\end{equation}
where $\omega_\alpha$ are harmonic $(1,1)$ forms and $\chi_i$ represent complex structure deformations; the displayed normalizations are schematic.
These inner products are precisely what determine the normalized fluctuations and hence the physical masses.
A stationary point that appears stable in coordinates $m^A$ is meaningful only after forming the covariant Hessian with the inverse field space metric, for example through blocks of the form
\begin{equation}
(M^2)^A{}_{B}\sim G^{A\bar C}\nabla_{\bar C}\partial_B V \,.
\end{equation}
Thus, numerical geometry is a diagnostic for false stabilization: flat or tachyonic directions can be hidden by poorly conditioned coordinates.

This observation is especially important for non-perturbative effects.
The real parts of K\"ahler moduli determine the exponential weights of worldsheet and brane instantons, while the imaginary parts are axions.
For instance, a contribution of the form $A\exp(-2\pi T_C)$ is exponentially sensitive to $T_C$, the complexified modulus whose real part is the calibrated volume of $C$. Yet the prefactor $A$ is not merely a constant in a realistic compactification.
It can contain one-loop determinants, Pfaffians, charged zero-mode dependence, and bundle moduli dependence; in heterotic compactifications these effects are central to whether a worldsheet instanton contributes to the superpotential at all~\cite{Witten:1999eg,Beasley:2003fx}.
A numerical Ricci-flat metric, combined with numerical Hermitian Yang--Mills data, gives a route to computing the spectra of the relevant Laplace and Dirac operators rather than treating these prefactors as unknowable order-one numbers.

D-term contributions provide another example.
They depend on slopes and Fayet--Iliopoulos terms, schematically
\begin{equation}
\xi \propto \frac{1}{\cV}\int_X J\wedge J\wedge c_1(L) \,,
\end{equation}
for a line bundle $L$.
The leading slope is cohomological, but deciding how the D-term fits into the full potential requires the gauge kinetic function, the charged matter K\"ahler metric, and the Hermitian Yang--Mills connection.
In other words, the same line bundle calculation that identifies a wall of stability is only the beginning; near the wall, canonical normalization and mixing with bundle moduli decide which fields become heavy, which are eaten, and which remain in the low-energy theory. The Fayet--Iliopoulos expression above is leading order and schematic, with possible changes from normalization conventions, loops, and higher-derivative corrections.

For type II and F-theory flux compactifications the same point becomes even sharper.
There one often writes
\begin{equation}
V = e^K\big(K^{A\bar B}D_AW D_{\bar B}\overline W - 3|W|^2\big) + V_D + \cdots \,,
\end{equation}
with the flux superpotential and warped backgrounds giving powerful mechanisms for fixing complex structure moduli and the axio-dilaton~\cite{Gukov:1999ya,Giddings:2001yu}.
KKLT and large volume scenarios then add non-perturbative and perturbative effects to address K\"ahler moduli~\cite{Kachru:2003aw,Balasubramanian:2005zx}.
But the practical questions --- mass hierarchies, tunneling directions, slow-roll parameters, whether a candidate minimum is really metastable --- depend on the metric on field space and on accurately computed cycle volumes.
A Ricci-flat metric also lets one check control conditions directly: the distribution of curvature in string units, the presence of localized small cycles, the separation between moduli masses and Kaluza--Klein scales, and the size of corrections that could invalidate the four-dimensional description.

\subsubsection{Open problems}

The most important stabilization problems that numerical metrics can sharpen are:
\begin{enumerate}[label=(\roman*)]
  \item \textbf{Complete heterotic stabilization in explicit realistic vacua:}
  One wants compactifications with a realistic visible sector and all geometric, bundle, and axionic moduli fixed in a regime of control.
  Bundle holomorphy stabilization of complex structure is known in principle~\cite{Anderson:2010mh}; the next step is precision stabilization with the actual K\"ahler metric, actual divisor volumes, and actual instanton actions.
  \item \textbf{Numerical Hessians and metastability:}
  Stabilization discussions often stop at parametric scaling.
  Numerical geometry allows one to compute the Hessian in canonically normalized fields and identify which directions are truly stabilized and which only appear stabilized in poor coordinates.
  \item \textbf{Instanton prefactors and dominant sectors:}
  The exponentials are controlled by cycle volumes, but deciding which instantons dominate also depends on zero modes and prefactors.
  Numerical metric data combined with spectral calculations should make this a computational problem rather than a qualitative guess.
  \item \textbf{Bundle metrics and charged matter normalization:}
  Heterotic stabilization depends on the gauge bundle as much as on the base geometry.
  Numerical Hermitian Yang--Mills connections, matter wavefunctions, and overlap integrals are needed to determine which bundle moduli are lifted and how strongly they mix with geometric moduli.
  \item \textbf{Control of the effective theory:}
  A candidate minimum should be checked against local curvature, small cycle limits, warping, and the Kaluza--Klein spectrum.
  A Ricci-flat metric supplies the data needed to decide whether the vacuum is inside the controlled supergravity regime or merely inside the formal K\"ahler cone.
  \item \textbf{Moduli dependence across families:}
  A single metric at a single point is not enough.
  One wants to track $\num{g}(m,\bar m)$ across moduli space, identify walls of stability, singular loci, and transitions between different EFT regimes.
\end{enumerate}

\subsection{Axions, inflation, and dark sectors}
\label{s:axindark}
A natural extension of the stabilization program is precision cosmology.
The questions we encounter sit at the intersection of compactification geometry, numerical differential geometry, swampland constraints, and observational cosmology.
In string compactifications, axions arise by expanding higher-form gauge potentials along internal harmonic forms, so their multiplicity is largely topological, while their physical normalization is geometric~\cite{Svrcek:2006yi,Arvanitaki:2009fg,Marsh:2015xka}.
The relevant four-dimensional kinetic terms are not fixed by intersection numbers alone: they require the Ricci-flat Calabi--Yau metric, since this metric defines the Hodge star and hence the $L^2$ pairing on harmonic forms.
Schematically, when an antisymmetric tensor field is expanded along harmonic two-forms $\{\omega_a\}$, one obtains kinetic matrices of the form
\begin{equation}
K_{ab} \sim \frac{1}{\cV}\int_X \omega_a\wedge *_g\,\omega_b ~,
\end{equation}
with analogous formulas in other duality frames.
Numerical Ricci-flat metrics therefore turn formal compactification data into canonically normalized fields and couplings.

This makes the metric directly relevant to alignment mechanisms, kinetic mixing, decay constants, reheating couplings, dark radiation, and the question of whether an explicit compactification genuinely realizes a desired inflationary or quintessence regime.
A large coordinate displacement in an axion basis need not remain large after diagonalizing $K_{ab}$, and a hierarchy in cycle volumes can be softened or enhanced by the full metric.
The same data enter instanton actions, gauge kinetic functions, overlap integrals, and matter couplings, so axion masses, photon couplings, hidden sector couplings, and reheating branching fractions are correlated rather than independent. This is especially important for multi-axion models such as $N$-flation and alignment scenarios~\cite{Dimopoulos:2005ac,Kim:2004rp}, for type IIB axiverse constructions~\cite{Cicoli:2012sz}, and for large ensembles where topology can generate many light fields~\cite{Demirtas:2018akl}.

The conceptual point is that a Ricci-flat metric provides the bridge between topology and cosmological observables.
It lets one ask whether an inflationary trajectory is long after canonical normalization, whether heavy moduli remain stabilized along that trajectory, whether instanton corrections respect the assumed hierarchy, and whether hidden sectors receive enough post-inflationary energy to be visible as dark radiation or too much energy to be viable.
These questions are inseparable from swampland constraints on long distance motion and quantum gravity consistency~\cite{Arkani-Hamed:2006emk,Ooguri:2006in}.

\subsubsection{Open problems}
The open problems here are both phenomenological and conceptual:
\begin{enumerate}[label=(\roman*)]
\item \textbf{Realistic axion sectors in explicit compactifications:}
Determine which explicit compactifications produce axion sectors with realistic decay constants and controllable kinetic and mass mixing matrices.

\item \textbf{Controlled large effective field ranges:}
Determine whether large effective field ranges can arise without leaving the regime of control, destabilizing moduli, or encountering unsuppressed instantons.

\item \textbf{Geometry of axion couplings and reheating:}
Compute how the same geometric data that generate hierarchical Yukawa couplings also shape axion couplings, reheating channels, and hidden sector temperatures.

\item \textbf{Exact multi-field inflationary kinematics:}
Compute covariant inflationary slow-roll parameters using the exact numerical field space metric to determine physical distances, rather than relying on asymptotic toy models, and determine when the resulting trajectories remain compatible with moduli stabilization.

\item \textbf{Compactification level dark sector signatures:}
Identify which darksector signatures, for example dark radiation, ultralight axion dark matter, birefringence, or fifth-force effects, are robust consequences of a compactification rather than adjustable inputs.
\end{enumerate}

\section{Mathematical aspects of the Calabi--Yau metric}
\label{s:maths}
This section collects the mathematical programmes that become accessible once an explicit Ricci-flat metric is in hand: the SYZ conjecture and special Lagrangian geometry, the search for analytic or semi-analytic metric expressions, restrictions of the total-space metric to fibrations, precision tests of mirror symmetry, and further directions in string mathematics.

\subsection{The SYZ conjecture and special Lagrangian geometry}
\label{s:SYZ}
Near appropriate large complex structure or large volume degenerations, the Strominger--Yau--Zaslow (SYZ) conjecture proposes that mirror pairs should admit dual special Lagrangian $T^3$-fibrations~\cite{Strominger:1996it,Kontsevich:2000hb}.
This is not merely a topological or Hodge theoretic slogan; in those degenerating regimes it is a metric expectation about the Ricci-flat representative in a degenerating K\"ahler class or complex structure family.
A submanifold $L\subset X$ is special Lagrangian of phase $\theta$ when
\begin{equation}
J|_L = 0, \qquad \Im(e^{-i\theta}\Omega)|_L = 0 ~,
\end{equation}
and $\Re(e^{-i\theta}\Omega)$ calibrates $L$ in the sense of Harvey--Lawson calibrated geometry~\cite{Harvey:1982xk}.
This calibration statement uses the orientation determined by the real part of the phase; with the opposite orientation one should compare the absolute calibrated volume.
The deformation theory is also metric sensitive: McLean's theorem identifies the infinitesimal deformations of a compact special Lagrangian with harmonic one-forms on $L$~\cite{McLean:1998ub}.
Thus a numerical Ricci-flat metric is exactly the object needed to turn the SYZ picture from an existence conjecture into a computational problem involving explicit calibrated cycles, their moduli, and their singular limits.

A practical numerical strategy is to search over embeddings or immersions $\iota:M^3\to X$ and minimize a calibration defect functional such as
\begin{equation}
\cE(\iota,\theta)=
\frac{\|\iota^*J\|^2}{\Vol(M)}+
\frac{\|\iota^*\Im(e^{-i\theta}\Omega)\|^2}{\Vol(M)}
+\lambda\,\bigg|\frac{\int_M\iota^*\Re(e^{-i\theta}\Omega)}{\Vol(M)}-1\bigg|^2 ~.
\end{equation}
The first two terms test the Lagrangian and phase conditions; the last term discourages collapse and fixes the calibrated normalization.
This optimization problem is difficult because the space of three-cycles is not given to us in advance.
Nevertheless, one can use several sources of structure: known homology classes, anti-holomorphic involutions and their real loci, torus actions in ambient toric descriptions, gradient flows of moment map coordinates, and asymptotic information near large complex structure.
The exact Ricci-flat metric then supplies the local volume form, Hodge star, and curvature data needed to decide whether a candidate is a genuine calibrated cycle or only an artifact of an algebraic approximation.

The singular fibers and the base affine structure are just as important as the smooth tori.
Away from the discriminant locus, dualizing the torus fibers gives the semi-flat mirror, but near the discriminant this construction misses instanton corrections from holomorphic discs with boundary on the fibers.
Joyce's picture of singular special Lagrangian fibrations, Gross's topological formulation of SYZ, and the Gross--Siebert program all emphasize that singular affine geometry is the organizing object~\cite{Joyce:2000js,Gross:1999tm,Gross:2007pf}.
Numerical metrics can test this structure directly: one can follow curvature concentration, measure the collapse of fiber diameters, reconstruct candidate affine coordinates on the base using semi-flat, McLean, or tropical data where available, and compare the observed behavior with the proven K3 case and with the best available threefold asymptotics~\cite{Gross:2000wp,Li:2019oyj,li2022strominger,Li:2022kqp,Li:2025mjm}.
For physics, the same data control A-brane moduli, disc instanton superpotentials, and D-brane central charges; for mathematics, they provide an experimental laboratory for the singularity theory of special Lagrangian geometry.

\subsubsection{Open problems}
The numerical SYZ program naturally leads to the following open questions:
\begin{enumerate}[label=(\roman*)]
  \item \textbf{Finding the right search space for three-cycles:}
  Can one build a systematic, non-\emph{ad hoc} parameterization of candidate three-cycles in compact \CY threefolds, perhaps combining homology, tropical data, ambient toric geometry, and learned embeddings?
  \item \textbf{Constructing compact special Lagrangian submanifolds numerically:}
  Can explicit compact special Lagrangian torus fibrations be found numerically on compact \CY threefolds, and can the resulting maps be made stable under refinement of the Ricci-flat metric?
  Using genetic algorithms,~\cite{AQR2026} recovers the known analytic examples: $\mathbb{RP}^3$ in the Fermat quintic and $T^3$ in the large complex structure limit of the Dwork family of quintics as special Lagrangian submanifolds of a Calabi--Yau threefold.
  It would be interesting to find other solutions.
  \item \textbf{Discriminant and singular fiber structure:}
  What singular fibers actually occur in generic compact examples, and how do they compare with Joyce's piecewise smooth models and with the affine structures predicted by Gross--Siebert theory~\cite{Joyce:2000js,Gross:2007pf}?
  \item \textbf{Semi-flat versus exact metrics:}
  Near large complex structure, can one measure the pointwise and spectral error between the exact numerical metric and the semi-flat approximation, including the rate at which the error concentrates near the discriminant?
  \item \textbf{Disc instanton corrections:}
  Can holomorphic discs ending on numerical special Lagrangian fibers be counted or approximated well enough to reproduce the wall crossing corrections of the SYZ mirror construction~\cite{Auroux:2009hf}?
  \item \textbf{Certified calibrated geometry:}
  Can interval arithmetic, a posteriori elliptic estimates, or Newton--Kantorovich methods turn an approximate numerical special Lagrangian into a rigorous existence theorem?
\end{enumerate}

\subsection{Analytic Ricci-flat Calabi--Yau metrics}
\label{s:anaCY}
Despite Yau's existence theorem, closed form analytic expressions for Ricci-flat K\"ahler metrics on compact Calabi--Yau manifolds are essentially unknown beyond flat tori and their finite orbifolds.
On the quintic threefold, on K3 realized as a complete intersection or Kummer surface, or on any smooth compact \CY with non-trivial cohomology that arises generically in string compactifications, no exact analytic metric is known in algebraic coordinates.
This is not merely a technical gap; it reflects the general difficulty of the compact complex Monge--Amp\`ere equation on K\"ahler manifolds with trivial canonical class, where existence is guaranteed but explicit integration by quadrature is not expected.

Between exact analytic metrics and purely numerical tables, several intermediate approaches produce partially explicit structures.

\paragraph{Analytic constructions on K3:}
For K3 surfaces, the hyper-K\"ahler structure provides additional algebraic rigidity that makes explicit metric constructions possible.
Kachru, Tripathy, and Zimet showed that BPS degeneracies of a heterotic little string theory compactified on $T^2$, assembled via D-geometry --- a formalism closely related to hyper-K\"ahler quotient constructions --- give a systematic construction of the Ricci-flat metric on smooth K3 surfaces~\cite{Kachru:2018lst,Kachru:2020ktm}.
For compact Calabi--Yau threefolds no analogous construction is available. Local metrics and asymptotic models exist in special non-compact settings, such as the conifold~\cite{Candelas:1989js,Stenzel:1993rf}, but no general perturbative expansion for compact threefold metrics around orbifold or conifold loci has been established, and the hyper-K\"ahler rigidity that makes the K3 case tractable is absent.

\paragraph{Machine learned algebraic Ansatz:}
Machine learning approaches produce K\"ahler potentials expressed as explicit functions of homogeneous coordinates: weighted sums of monomials in the balanced metric formalism, or neural network outputs trained on the Monge--Amp\`ere residual~\cite{Ashmore:2019wzb,Jejjala:2020wcc,Douglas:2020hpv,Douglas:2021zdn,Berglund:2022gvm,Berglund:2024psp}.
These representations are concrete and evaluable --- they are not just tables --- and in this sense constitute semi-analytic expressions for the metric, even if not closed form in the classical sense.
They have been applied to compute physical Yukawa couplings and matter K\"ahler metrics in explicit heterotic models~\cite{Butbaia:2024tje,Berglund:2024uqv,Constantin:2024yxh}, demonstrating that semi-analytic representations are sufficient for quantitative phenomenology.

\paragraph{Analytic and numerical studies of complex structure dependence:}
A decisive recent development is the extension of numerical \CY metrics from isolated points to \emph{families} parametrized by complex structure moduli.
Anderson et al.\ carried out the first systematic computation of moduli dependent \CY metrics, using neural networks trained to output the metric as a function of complex structure parameters --- a purely numerical approach --- giving the first quantitative picture of how the metric deforms continuously across moduli space~\cite{Anderson:2020hux,Larfors:2021pbb}.
A complementary analytic perspective has been opened by Mirjani\'c and Mishra, who exploited the extrinsic symmetry group of Calabi--Yau hypersurfaces to derive compact symbolic approximations to K\"ahler metrics on specific loci of the Dwork one-parameter family and the Fermat family, show that symmetry strongly constrains the metric on those loci, and distill machine learned models into compact algebraic expressions~\cite{Mirjanic:2024srm}.
Together these results give a first analytic and numerical handle on how the Ricci-flat metric varies within families of compact \CY manifolds, while still requiring residual/error checks rather than claiming exact closed-form metrics.

Along the same line, Lee and Lukas~\cite{Lee:2025pue} have recently proposed a direct route from numerical Ricci-flat metrics to compact analytic formul\ae.
Starting with machine learned approximations to the Ricci-flat K\"ahler potential, they fit the resulting data to low-degree Donaldson Ansatz, with the large discrete symmetry of the family used to reduce the number of allowed projective invariants.
For the Dwork one-parameter family of quintics in $\mathbb P^4$ and an analogous one-parameter family of bi-cubic hypersurfaces in $\mathbb P^2\times \mathbb P^2$, this procedure yields simple explicit K\"ahler potentials whose dependence on the complex structure parameter is captured by a small number of fitted functions, apparently depending only on the modulus of that parameter.
The resulting metrics are not exact solutions of the compact Monge--Amp\`ere equation, and the approximation is least reliable near high-curvature regions such as the conifold locus, but the construction is an important proof of principle: neural numerical data can be distilled into explicit, symmetry controlled, semi-analytic K\"ahler potentials that are accurate at the percent to few percent level and are therefore suitable for testing metric dependent quantities beyond what period data alone can determine.

These developments enable metric level studies of phenomena inaccessible from period integrals alone:
\begin{itemize}
  \item how physical Yukawa couplings, canonical normalizations, and matter K\"ahler metrics vary under complex structure deformations, and whether approximate flavor symmetries are preserved or lifted across moduli space;
  \item the behavior of the metric approaching singular loci such as conifold transitions, where a three-cycle collapses, curvature concentrates, and the effective four-dimensional theory develops new light states;
  \item the accuracy of naive approximations --- product Ansatz, adiabatic fibration, moduli factorization --- as a function of position in moduli space, providing quantitative bounds rather than parametric estimates.
\end{itemize}

The moduli dependent metric is the enabling object for a number of the physics programs described in preceding sections.
Precision moduli stabilization requires the K\"ahler metric on the space of complex structure deformations; swampland distance conjecture tests require geodesic distances in field space; and precision mirror symmetry tests require that metric quantities on both sides be computed in the same moduli coordinates.
All of these ultimately require not one metric but a continuously varying family.

\paragraph{The large complex structure limit and semi-flat approximation:}
The most analytically controlled regime in which compact \CY metrics can be approximated explicitly is the large complex structure (LCS) limit.
As the complex structure approaches this boundary of moduli space, the \CY is expected to develop a special Lagrangian $T^3$-fibration with flat fiber metrics, and instanton corrections are exponentially suppressed in the fiber area --- the picture mandated by the SYZ conjecture~\cite{Strominger:1996it}.

The leading order metric in this limit is the \emph{semi-flat} metric: a Ricci-flat K\"ahler metric on the total space of the fibration that restricts to a flat metric on each smooth torus fiber and whose base direction part is determined by a real Monge--Amp\`ere equation on the affine base.
For K3 surfaces, Gross and Wilson proved convergence of the exact Yau metrics to the semi-flat metric away from the discriminant, together with a detailed description of the collapsing limit near singular fibers~\cite{Gross:2000wp}.
For higher-dimensional compact \CY manifolds, including threefolds, the situation is more subtle. Li's foundational work establishes metric SYZ convergence for the Dwork one-parameter family of quintic threefolds, conditional on a non-Archimedean conjecture, and obtains asymptotics near discriminant loci in related families~\cite{Li:2020syz,Li:2022kqp}. Boucksom--Jonsson~\cite{Boucksom:2016khs} and Tosatti~\cite{Tosatti:2020nyp} develop the non-Archimedean framework. More recently, Li extended the estimates to the \emph{intermediate} complex structure regime (away from the large complex structure point)~\cite{Li:2025mjm},
and with Tosatti established generic regularity of such limits~\cite{LiTosatti:2025}; a companion paper treats degenerations whose essential skeleton has intermediate
dimension~\cite{Li:2025deg}. Two very recent papers significantly advance the general programme: Blum--Liu~\cite{Blum:2026viy} construct \emph{valuatively independent} canonical bases for sections of an ample line bundle on log \CY pairs over a non-Archimedean field --- the combinatorial key needed to control the $C^0$-limit of the K\"ahler potential ---
and Li~\cite{Li:2026syz} proves that any polarised maximal degeneration admitting a valuatively independent canonical basis satisfies the metric SYZ conjecture. Together these results constitute the sharpest available bridge between Yau's theorem and explicit asymptotic metric formul\ae\ for log \CY degenerations; an unconditional general theorem for all compact \CY threefold degenerations remains open.
Numerically, one can test and quantify the quality of the semi-flat approximation by computing the exact Ricci-flat metric at large but finite complex structure and comparing it pointwise with the semi-flat formula.
This approach sheds light on:
\begin{itemize}[topsep=-2pt, itemsep=-1pt]
  \item the rate of convergence to the SYZ picture as a function of distance from the LCS point in moduli space, and hence the effective size of disc instanton corrections;
  \item the structure of the discriminant locus, where the $T^3$ fibers degenerate and singular fiber types can in principle be read off from curvature concentration;
  \item the affine structure on the base of the fibration, which is the central input to the Gross--Siebert program for a symplectic geometry reconstruction of the mirror~\cite{Gross:1999tm}.
\end{itemize}

Numerical \CY geometry thus provides an independent, metric level laboratory for the SYZ conjecture: one can observe the emergence of special Lagrangian fibrations, measure the affine base structure, and directly compare instanton corrections with predictions from genus-zero Gromov--Witten invariants on the mirror.
This role is complementary to, and largely independent of, the period based tests of mirror symmetry discussed elsewhere in this note.

\subsubsection{Open problems}

\begin{enumerate}[label=(\roman*)]
\item \textbf{Analytic or asymptotic formulas beyond flat tori:}
Can any rigorous analytic or asymptotic expression for a compact \CY metric be established beyond the flat torus baseline, for instance as a controlled expansion around an orbifold or conifold locus, or in the LCS limit?

\item \textbf{Instanton corrections to the semi-flat approximation:}
Can the semi-flat approximation on compact \CY threefolds be systematically corrected by disc instanton contributions, and can those corrections be computed numerically and compared with mirror Gromov--Witten invariants?

\item \textbf{Compression of machine learning Ansatz:}
Can machine learning Ansatz be compressed into compact algebraic or rational expressions --- replacing high dimensional network weights by structured functional forms --- while maintaining useful accuracy?

\item \textbf{Metric families over multi-parameter moduli spaces:}
Can moduli dependent metric families be computed efficiently across multi-parameter complex structure spaces, enabling metric level landscape exploration at a cost polynomial rather than exponential in the number of moduli?
\end{enumerate}

\subsection{Restrictions of the Ricci-flat metric}
\label{s:restrict}
Many Calabi--Yau manifolds appearing in string compactification literature are fibered, and fibrations are especially common in large classes of toric hypersurfaces and complete intersections~\cite{Avram:1996xn,Candelas:2012az,Anderson:2017aux,Huang:2018esr}.  Given an explicit, and in cases of interest here --- numerically computed Ricci-flat metric on the total space, a natural question is what metric this induces on a generic fiber and on the base.

The restriction of the total space metric to a smooth fiber is not, in general, the intrinsic Ricci-flat metric in the induced K\"ahler class on that fiber.  It is the first fundamental form (the induced Riemannian metric) inherited from the embedding, and it generally contains extrinsic curvature corrections.  In an adiabatic limit, where the fiber volume is small compared with the base scale, the restricted metric may approach a semi-flat or fiberwise Ricci-flat approximation, but this is an additional limiting assumption rather than an exact identity. Reverse engineering this assumption then provides a perturbative framework to compute the induced metric on the fiber from a Ricci-flat metric on the total space.

Similarly, a metric on the base is obtained only after choosing a horizontal distribution, for example the $g$-orthogonal complement to the vertical tangent bundle.  The resulting Riemannian-submersion metric depends on the fibration, possible singular fibers (the locations and severity of singularization, etc.), and the scale separation.  Near discriminant loci one should expect corrections and singular behavior rather than a smooth Ricci-flat base metric; we will revisit this concept from a different perspective in \S\,\ref{s:cosmic}.

This question is closely related to earlier degeneration and singularization problems: as fibers collapse, numerical metrics can track curvature concentration, monodromy, and the approach to local models.  It is also related to Tyurin degenerations, where a Calabi--Yau degenerates into quasi-Fano components glued along a common anticanonical Calabi--Yau divisor --- possibly even a smooth submanifold~\cite{tyurin2003fano,Doran:2016uew,Kanazawa:2016tnt,Berglund:2022dgb}.  A numerical metric could make the neck region and limiting geometry explicit.

\subsubsection{Open problems}

\begin{enumerate}[label=(\roman*)]
  \item \textbf{Intrinsic Ricci-flatness of fibers:}
  Can one quantify the failure of the restricted metric on a generic fiber to be intrinsically Ricci-flat, and compare it with the fiber's Yau metric in the induced K\"ahler class?

  \item \textbf{Horizontal base metric:}
  Can one construct the base metric using a horizontal distribution and compare it with semi-flat or adiabatic predictions?

  \item \textbf{Singular fibers in the total space metric:}
  How do singular fibers appear in the numerical Ricci-flat metric on the total space?

  \item \textbf{Tyurin degenerations and mirror decompositions:}
  Can Ricci-flat metrics be used to test Tyurin degeneration pictures and other decompositions used in mirror symmetry?
\end{enumerate}

\subsection{Precision tests of mirror symmetry}
\label{s:precMS}
Much of mirror symmetry is traditionally tested on protected quantities: Hodge numbers, period integrals, genus-zero prepotentials, Gromov--Witten invariants, monodromy data, and the enumerative predictions first made concrete for the quintic threefold~\cite{Candelas:1990rm}.
These tests are profound, but they largely avoid the non-holomorphic information that enters physical normalization.
Numerical Ricci-flat metrics make a more ambitious program possible: compute a metric dependent observable directly on one side, compute the mirror prediction using periods, instanton expansions, or special geometry on the other side, and compare the two with independent error bars~\cite{Demirtas:2023als,Butbaia:2024tje}.

The cleanest starting point is the Weil--Petersson metric.
On the complex structure side it is obtained from
\begin{equation}
K_\text{cs}=-\log\bigg(i\int_X\Omega\wedge\overline\Omega\bigg) ~,
\qquad
G^\text{cs}_{a\bar b}=\partial_a\partial_{\bar b}K_\text{cs} ~,
\end{equation}
where periods give a highly constrained B-model computation.
On the mirror side, the same object should be reproduced by direct numerical integration of harmonic representatives and the Hodge star of the Ricci-flat metric. On the A-model/K\"ahler side, such a comparison should use the quantum-corrected K\"ahler metric, or be restricted to large radius where worldsheet instantons are negligible.
This comparison is valuable because the two calculations fail in different ways: period calculations are sensitive to analytic continuation, monodromy, truncation of instanton series, and coordinate choices, while metric calculations are sensitive to sampling, Ansatz error, harmonic projection, and quadrature.
Agreement would therefore be a strong certification of both methods.

The same logic applies to normalized Yukawa couplings and brane data.
Holomorphic Yukawa couplings can often be obtained algebraically, but physical Yukawa couplings require matter metrics and canonical normalization.
Likewise, open string mirror symmetry predicts brane superpotentials, central charges, and wall crossing behavior, while a direct A-model calculation requires special Lagrangians, flat bundles, and disc instantons~\cite{Auroux:2009hf,Abouzaid:2012kk}.
Numerical metrics supply the missing geometric inputs: harmonic forms, calibrated volumes, Laplacian spectra, and the local data needed for overlap integrals.
This changes mirror symmetry from a comparison of formal expansions into a precision measurement problem.

A particularly important target is the boundary of moduli space.
Large complex structure, conifold, and Landau--Ginzburg points come with asymptotic formulas, but a compactification used in physics usually sits at finite distance from those limits.
With moduli dependent metrics one can ask how far into the interior a given asymptotic expansion remains accurate, how canonical normalization changes near a conifold, and whether towers of light states inferred from periods are visible in the numerical Kaluza--Klein spectrum~\cite{Ashmore:2021qdf}.
This gives a way to test mirror symmetry at the level of actual scales and couplings rather than only at the level of protected holomorphic data.

\subsubsection{Open problems}
Several precision tests now look feasible:
\begin{enumerate}[label=(\roman*)]
  \item \textbf{Multi-parameter mirror pairs:}
  Can direct numerical special geometry be compared with mirror period computations in compact examples with large $h^{2,1}$ or large $h^{1,1}$, where analytic continuation and instanton summation are genuinely high dimensional?
  \item \textbf{Canonical normalization of Yukawa couplings:}
  Can normalized Yukawa couplings, not just holomorphic Yukawa couplings, be computed on both sides of mirror symmetry with controlled uncertainties?
  \item \textbf{Open string mirror symmetry:}
  Can numerical special Lagrangians, brane moduli metrics, disc instantons, and central charges be computed accurately enough to test compact open string mirror symmetry?
  \item \textbf{Asymptotic error maps:}
  Can one produce quantitative maps over moduli space showing where large complex structure, conifold, or Landau--Ginzburg approximations are reliable and where they fail?
  \item \textbf{Topology change and phase structure:}
  Can canonically normalized quantities be tracked through extremal transitions, flop walls, and hybrid phases, and then compared with the mirror description of the same path?
  \item \textbf{Independent error budgets:}
  Can the community develop benchmark mirror pairs for which period truncation error, numerical Ricci-flat error, harmonic representative error, and quadrature error are all reported in compatible conventions?
\end{enumerate}

\subsection{Further directions in string mathematics}
\label{s:further}
The mathematical uses of numerical Ricci-flat metrics are not secondary; they are one of the main reasons this technology matters.
A compact \CY metric is a solution of a nonlinear elliptic problem whose explicit form is almost never known.
Once approximated with controlled error, it becomes a laboratory for questions in complex differential geometry, geometric analysis, calibrated geometry, and topology.
This is analogous to the role of numerical experiments in number theory or low-dimensional topology: the computation does not replace proof, but it can reveal the correct conjecture, identify exceptional behavior, and provide initial data for rigorous arguments.

One important direction is numerical Hermitian Yang--Mills theory.
Generalized Donaldson algorithms and their descendants approximate HYM connections on explicit holomorphic bundles~\cite{Anderson:2010ke}.
Combined with a Ricci-flat background metric, these methods let one study how bundle metrics behave near stability walls, distinguish wall crossing from actual singular-sheaf limits, analyze possible Uhlenbeck bubbling or curvature concentration, and determine how matter zero modes localize.
The resulting data are mathematically interesting in their own right because they probe the analytic meaning of slope stability and the limiting behavior predicted by the Donaldson--Uhlenbeck--Yau correspondence~\cite{Donaldson:1985zz,Uhlenbeck:1986de}.
They are also the input for heterotic effective field theory.

A second direction is spectral geometry.
The scalar, $(p,q)$-form, Dolbeault, and Dirac spectra of compact \CY manifolds are almost unexplored in explicit threefold examples.
Numerical metrics make it possible to compute spectral gaps, multiplicities, eigenfunction localization, heat kernel coefficients, analytic torsion, and the variation of these quantities over moduli space.
These spectra are central to Kaluza--Klein reduction and threshold corrections, but they also raise pure mathematical questions: how much of the complex or K\"ahler geometry is audible, whether mirror pairs have related spectral statistics, and how spectra behave under degeneration.
Existing balanced metric and energy functional methods provide the geometric starting point for such calculations~\cite{Headrick:2005ch,Donaldson:2005hvr,Douglas:2006rr,Headrick:2009jz}.

Metric degeneration and topology change form a third arena.
Conifolds, extremal transitions, collapsing fibrations, and Gromov--Hausdorff limits are central both in string theory and in modern geometry.
Numerical metrics allow one to watch curvature concentrate, cycles collapse, eigenvalues move, and harmonic forms change support.
In the large complex structure regime this connects directly to SYZ collapse and to rigorous results for K3 and conditional results for higher-dimensional \CY manifolds~\cite{Gross:2000wp,Li:2019oyj,li2022strominger,Li:2022kqp,Li:2025mjm}.
Near conifold points it can reveal how local models such as the deformed or resolved conifold are glued into a compact metric.

Ricci-flat metrics should also be useful for Reid's conjecture~\cite{Reid:1987}, not because they by themselves prove the existence of the required algebraic transitions, but because they make the proposed connected web of \CY threefolds geometrically testable.
Reid's picture suggests that apparently disconnected families of \CY threefolds may be related by passing through singular spaces, especially conifold transitions, and the physics literature has long treated such transitions as genuine paths between compactification geometries~\cite{Green:1988bp,Green:1988wa,Candelas:1989ug,Candelas:1989qn,Avram:1995pu,Avram:1997rs}; see also~\cite{Anderson:2022bpo}.
A numerical metric gives local and global diagnostics for these passages: one can watch exceptional curves or vanishing three-spheres collapse, measure curvature concentration near ordinary double points, compare the compact metric with local deformed and resolved conifold models, and ask whether the two sides approach a common metric limit~\cite{Candelas:1989js}.
This would turn Reid's conjecture from a primarily topological and algebro-geometric connectedness proposal into a quantitative program in metric geometry: map candidate transition paths, identify which singular limits are actually reached by controlled Ricci-flat metrics, and determine which physical quantities remain finite, jump, or degenerate as one moves through the web of \CY compactifications.

Finally, numerical metrics broaden the study of calibrated and special holonomy geometry.
Real loci of \CY manifolds fixed by anti-holomorphic isometric involutions give natural three-manifolds, sometimes with calibrated or nearly calibrated structures, and recent work has begun computing harmonic forms on such loci using numerical \CY metrics~\cite{Douglas:2024pmn}.
This suggests a path from explicit \CY data to explicit $G_2$ questions: associative and co-associative cycles, twisted connected sums, and analytic gluing constructions can all benefit from concrete metrics, spectra, and harmonic forms.
(In addition to~\cite{Douglas:2024pmn}, other machine learning works on metrics on $G_2$ manifolds include~\cite{Aggarwal:2023swe,Heyes:2026rch}.)
At the rigorous end of the spectrum, numerical solutions can seed Newton iterations, interval arithmetic estimates, and computer assisted proofs, turning experimental \CY geometry into certified geometry.

\subsubsection{Open problems}
A few mathematical problems stand out:
\begin{enumerate}[label=(\roman*)]
  \item \textbf{Numerical HYM degeneration:}
  Can one track HYM connections through stability walls and identify the analytic limit, including curvature concentration and the appearance of singular sheaves?
  \item \textbf{Spectral atlases of compact \CY manifolds:}
  Can benchmark datasets of scalar, form, Dolbeault, and Dirac spectra be produced for standard hypersurfaces, complete intersections, quotients, and their mirrors?
  \item \textbf{Spectral mirror symmetry:}
  Are there metric level spectral regularities shared by mirror pairs once volumes, moduli coordinates, and canonical normalizations are matched?
  \item \textbf{Degeneration diagnostics:}
  Which numerical observables best detect conifold formation, collapsing fibrations, Gromov--Hausdorff limits, or topology change before the algebraic model becomes singular?
  \item \textbf{Calibrated real loci and $G_2$ constructions:}
  Can numerical \CY metrics be used to find calibrated real loci and to supply gluing data for explicit special holonomy metrics in seven dimensions?
  \item \textbf{From numerics to proof:}
  Can a posteriori estimates, interval methods, and certified elliptic solvers prove existence or non-existence statements suggested by numerical Ricci-flat metrics?
\end{enumerate}

\section{Emergent field space and spacetime-variable geometries}

This section collects the directions in which an explicit Ricci-flat \CY metric enters questions about field space geometry and the spacetime interpretation of compactification data.
The first theme is the quantitative study of swampland distances, light towers, level crossing, and emergent field space metrics, where one wants to compare moduli space distances and physical spectra in the same compact model.
The second theme is more speculative but closely related: allowing the internal \CY geometry, its moduli, or its metric data to vary over the observable spacetime, as in stringy cosmic strings, spacetime-variable compactifications, and higher-dimensional geometries in which four-dimensional spacetime is not assumed to be a rigid product factor.
In both cases, the numerical metric is not the entire solution, but it supplies the local Hodge theory, Laplacians, kinetic terms, volumes, and curvature data needed to turn qualitative quantum gravity and spacetime geometry ideas into concrete calculations.

\subsection{Swampland distances, towers, and emergent field space geometry}
\label{s:swamp}
The swampland distance conjecture is, at its most operational, a metric statement about moduli space~\cite{Ooguri:2006in,Palti:2019pca}.
The field space distance along a path $\gamma$ is
\begin{equation}
\Theta(\gamma)=\int_\gamma \sqrt{G_{A\bar B}\,\dot\phi^A \dot{\bar\phi}^{\bar B}}\,\dd s ~,
\end{equation}
and the conjecture predicts an infinite tower of states whose mass scales as
\begin{equation}
m_\text{tower} \sim m_0 e^{-\lambda\Theta}
\end{equation}
in appropriate asymptotic limits.
To test this in a compact model one needs two ingredients which are rarely available together: the moduli space metric and the spectrum of physical masses over the same moduli.
A numerical Ricci-flat \CY metric is useful precisely because it connects these data.
It gives differential geometric control of the compactification and allows one to build operators --- most simply the scalar Laplacian --- whose eigenvalues determine the Kaluza--Klein scale.
This is the logic behind the numerical program developed in~\cite{Ashmore:2019wzb,Ashmore:2021qcb,Anderson:2020hux,Larfors:2022nep}.

The cleanest illustration is the quintic study of Ashmore and Ruehle~\cite{Ashmore:2021qdf}.
Numerical \CY metrics are used there to follow the low lying Kaluza--Klein spectrum over K\"ahler and complex structure moduli, while geodesic distances are computed using mirror symmetry, approximations, and direct numerical methods.
The comparison makes the distance conjecture quantitative: one can identify the eigenvalues defining the light tower, fit $m_\text{tower}(\Theta)$, and extract the exponential rate $\lambda$.
It also shows that the tower should be treated as a spectral object rather than as a fixed label attached to a single mode.
Level crossing can occur, and the lightest states may be selected by global features of the compact space, not only by a na\"{\i}ve radius.
Level crossings in the scalar Laplacian spectrum, computed over complex structure moduli for tori, the quartic K3, and the quintic threefold, are related to complex multiplication and rank-one attractor points~\cite{Ahmed:2023cnw}.

This makes several previously vague questions sharp.
Which tower actually becomes light near a given infinite distance limit: Kaluza--Klein modes, wrapped branes, string excitations, or a mixture whose identity changes along the path? 
How stable is $\lambda$ under changes of path, normalization, and finite distance corrections?
How does the story change in multi-parameter moduli spaces or across birational phases?
Ruehle and collaborators have emphasized these global issues in their work on infinite distance limits, extended K\"ahler cones, and infinite flop chains~\cite{Grimm:2019ixq,Brodie:2021toe,Brodie:2021xsf}.
Limiting mixed Hodge structures and intersection data identify possible boundary regimes; numerical geodesics and spectra can test how quickly the asymptotic behavior sets in.

The heterotic case adds another layer.
The light fields need not be only the familiar K\"ahler and complex structure moduli of the underlying \CY threefold; Hermitian, bundle, and gauge degrees of freedom can also participate.
The heterotic K\"ahler gravity analysis connects large distance behavior to an effective holomorphic theory for complexified Hermitian and gauge data~\cite{MurgasIbarra:2024lks}.
A genuinely heterotic test of the distance conjecture should therefore combine numerical Ricci-flat metrics, approximate Hermitian-Yang--Mills data, and four-dimensional kinetic terms.

An especially attractive open direction is to connect this with Yukawa couplings and canonical normalizations.
If field space geometry is partly emergent from integrating out towers of states~\cite{Heidenreich:2018kpg}, then the same numerical technology used to compute physical Yukawa couplings, kinetic terms, and massive spectra should also illuminate swampland behavior.
Here, phenomenology, quantum gravity conjectures, and numerical geometry point to the same calculation.

\subsubsection{Open problems}

\begin{enumerate}[label=(\roman*)]
\item \textbf{Distances and spectra in the same compact model:}
Can one compute the moduli space metric, geodesic distance, and tower spectrum over the same region of moduli space in explicit compact models, so that the exponential rate $\lambda$ in $m_{\mathrm{tower}}\sim m_0 e^{-\lambda\Theta}$ can be extracted with controlled numerical uncertainties?

\item \textbf{Identifying the light tower:}
Determine, in explicit degenerating families, whether the tower predicted by the distance conjecture is made from Kaluza--Klein modes, wrapped branes, string excitations, gauge or bundle modes, or mixtures whose identity changes along the path through level crossing.

\item \textbf{Finite distance corrections and onset of the asymptotic regime:}
Quantify how far from an infinite distance boundary the asymptotic exponential behavior becomes reliable, and determine which finite distance corrections control the approach to the limiting tower behavior.

\item \textbf{Multi-parameter moduli spaces and path dependence:}
Extend numerical tests of distance conjectures from one-parameter paths to genuine multi-parameter moduli spaces, including non-geodesic paths, monodromy orbits, flop walls, extended K\"ahler cones, and birational phase transitions.

\item \textbf{Heterotic distance conjectures with bundle data:}
Develop a genuinely heterotic test of infinite distance behavior that includes not only the Ricci-flat metric on $X$, but also numerical Hermitian Yang--Mills connections, bundle moduli, charged matter metrics, and the spectra of gauge and bundle-valued kinetic operators.

\item \textbf{Emergent field space geometry from towers:}
Test whether the numerically computed moduli space metric can be reconstructed, even approximately, from the tower of massive states and its threshold corrections, thereby making the emergent field space picture quantitative in compact examples.

\item \textbf{Swampland tests with full EFT control data:}
Combine distance, tower, and species-scale diagnostics with local curvature, warped scales, cycle volumes, and Kaluza--Klein gaps, so that apparent swampland behavior can be distinguished from loss of control in the underlying compactification.
\end{enumerate}

\subsection{Spacetime-variable Calabi--Yau compactifications}
\label{s:cosmic}
The foregoing discussion pertains to compactification Ans\"atze where the total, ten-dimen\-si\-sonal spacetime is assumed to have the rigid product structure, 
$M^{1,3}\times X$, with $M^{1,3}$ denoting the observable spacetime, while $X$ is a fixed, complex 3-dimensional, compact Calabi--Yau manifold, too small to be directly observable. Correspondingly, the metric is throughout assumed to be a direct sum of the spacetime metric on $M^{1,3}$ and the Ricci-flat/Calabi--Yau metric on $X$, completely independent of each other.
We now relax this rigid product structure and include some increasingly more speculative proposals.

\paragraph{Stringy Cosmic Strings:}
This general framework clearly admits generalizations, the simplest of which allows the Calabi--Yau manifold, $X$, to vary (be fibered nontrivially) over the spacetime, $M^{1,3}$, by making the moduli of $X$ vary over $M^{1,3}$. This gives rise to the configurations known as ``stringy cosmic strings,'' where
the $M^{1,3}$ metric is {\em\/warped,} and the moduli fields $\phi^i(x)$ are four-dimensional fields whose kinetic terms are governed by the Weil--Petersson or Zamolodchikov metric on moduli space~\cite{Greene:1989ya,Green:1993zr,Berglund:2000fy}.  After dimensional reduction one expects equations schematically of the general form
\begin{equation}
R_{\mu\nu}=G_{i\bar j}^{\rm WP/Z}(\phi)\,
 \partial_\mu \phi^i\,\partial_\nu \bar \phi^{\bar\jmath}+\cdots,
\end{equation}
where the ellipsis denotes warp factors, fluxes, localized sources, potential terms, and higher-derivative corrections when present.  Thereby, a numerical Ricci-flat metric can supply the local moduli space metric and the variation of internal volumes, but a full background still requires solving the coupled spacetime and internal equations.

If $X$ varies either only along, $S$, a spatial surface factor in spacetime, $(S{\times}\mathbb{R}^{1,1})=M^{1,3}$ (or along the entire $M^{1,3}$), it is shown that $X$ {\em\/must\/} singularize at some real codimension-2 locations in $M^{1,3}$, the Euler number of which is an $X$-dependent topological invariant~\cite{Greene:1989ya,Green:1993zr,Berglund:2000fy}.
In such codimension-two examples, variation around a defect is often described by a holomorphic map from the transverse plane into moduli space.  Compactifying the model by adding (temporarily) a limiting $X_\infty$-fiber at ``infinity'' of $M^{1,3}$, this can be related to degenerations of Calabi--Yau families over a base; near discriminant loci the numerical metric can probe how cycles collapse and how the effective spacetime stress tensor is distributed.

There is also a possible complementary non-compact viewpoint: If $X_\infty$ is an anticanonical (Calabi--Yau) divisor in a Fano manifold $F$, then under suitable hypotheses $F\smallsetminus X_\infty$ can carry complete Calabi--Yau metrics~\cite{Tian:1990ty,Tian:1991swq}. 
These metrics are governed by normalized or asymptotic Monge--Amp\`ere problems rather than by a naive equality of divergent global integrals.  Numerical approximations to them could provide local models for spacetime-dependent compactifications and for asymptotic regions of moduli space.
Whether the stringy cosmic string configuration of the previous two paragraphs, the total {\em non-compact\/} spacetime of the fibration, $S\rtimes X$ (or the entire $M^{1,3}\rtimes X$ with a Euclidean metric) can be identified metrically and to what degree/approximation with the Tian--Yau non-compact Calabi--Yau manifold $F\smallsetminus X_\infty$, is tantalizing but unknown. It however seems well worth exploring --- and becomes possible with explicit numerical metrics computable on both sides of this potential identification.

Note the conceptual complementarity with the discussion of fibrations in \S\,\ref{s:restrict}: There, one computes the numerical Ricci-flat metric on an elliptically or K3-fibered Calabi--Yau manifold, so as to compare the induced metric on the (lower-dimensional Calabi--Yau) fibers with their intrinsic Ricci-flat metric. Here, one computes the numerical Ricci-flat metric on a family of Calabi--Yau manifolds fibered over a spatial surface factor, $S$, or the entire spacetime, $M^{1,3}$, and can use the warped metric on $M^{1,3}$ to compute the metric on the total, fibered spacetime, $M^{1,3}\rtimes X$, which ought to itself be Ricci-flat in string theory. 
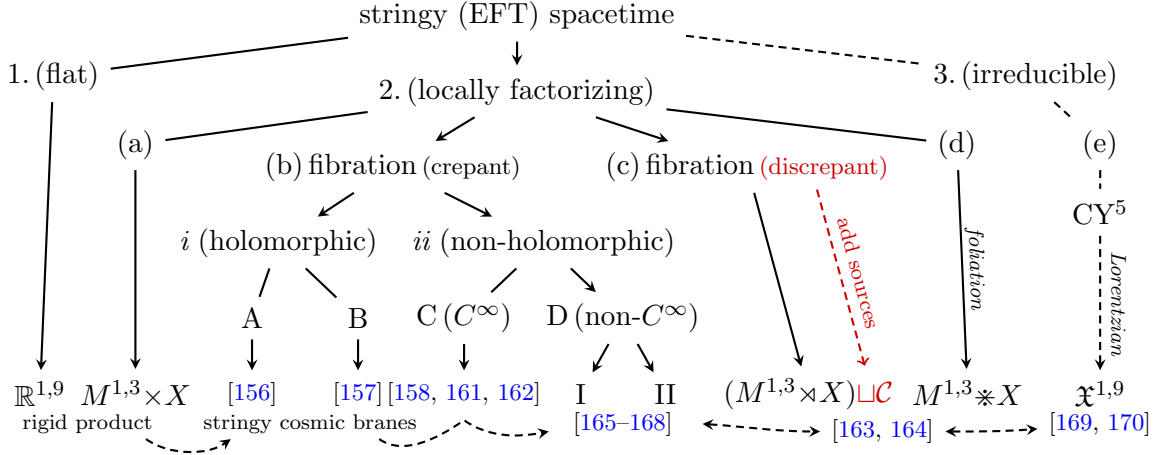
\begin{figure}[htb]
 \centering
   \begin{tikzpicture}[xscale=1.4,
        every node/.style={inner sep=1mm,outer sep=.5mm}]
     \path[use as bounding box](-.2,.4)--(10.2,-5.5);
     \node(S) at(4.5,0) {stringy (EFT) spacetime};
      \node(1) at(.125,-.8) {1.\,(flat)};
       \node(F) at(0,-5) {$\IR^{1,9}$};
      \node(2) at(4.5,-1) {2.\,(locally factorizing)};
       \node(2a) at(.9,-1.7) {(a)};
        \node(T) at(.9,-5) {$M^{1,3}{\times}X$};
       \node(2b) at(3.33,-2) {(b)\,fibration\,\footnotesize(crepant)};
        \node(2bi) at(2.25,-3) {{\it i} (holomorphic)};
         \node(2biA) at(2,-4) {A};
          \node(SCS) at(2,-5) {\footnotesize\cite{Greene:1989ya}};
         \node(2biB) at(3,-4) {B};
          \node(CYCY) at(3,-5) {\footnotesize\cite{Green:1993zr}};
        \node(2bii) at(4.75,-3) {{\it ii} (non-holomorphic)};
         \node(2biiA) at(4,-4) {C\,($C^\infty$)};
          \node(BHM1-3) at(4,-5) {\footnotesize\cite{Berglund:2000fy,Berglund:2000gt,Berglund:2000mr}};
         \node(2biiB) at(5.5,-4) {D\,(non-$C^\infty$)};
          \node(I) at(5.1,-5) {I};
          \node(II) at(5.9,-5) {II};
       \node(2c) at(6.67,-2) {(c)\,fibration\,\color{red!80!black}\footnotesize(discrepant)};
        \node(fd) at(7.25,-5)
                    {$(M^{1,3}{\rtimes}X)\color{red!80!black}{\sqcup}\mathcal{C}$};
        \draw[red!80!black, densely dashed, ->, thick](7.33,-2.3) to node
              [above, rotate=-75]{\scriptsize\color{red!85!black}add sources}
               ++(.45,-2.33);
       \node(2d) at(8.65,-1.7) {(d)};
        \node(fo) at(8.75,-5) {$M^{1,3}{\divideontimes}X$};
      \node(3) at(9.3,-.8) {3.\,(irreducible)};
       \node(3e) at(10,-1.7) {(e)};
       \node(5) at(10,-2.6) {CY$^5$};
       \node(HH) at(10,-5) {$\mathfrak{X}^{1,9}$};
       \node(A-dS) at(7.95,-5.5) {\footnotesize\cite{Banerjee:2018qey,Blaback:2019zig}};
     \draw[-stealth, thick](S)--(1)--(F);
     \draw[-stealth, thick](S)--(2); \draw[-stealth,thick](2)--(2a)--(T);
      \draw[-stealth, thick](2)--(2b);
       \draw[-stealth, thick](2b)--(2bi);
        \draw[-stealth, thick](2bi)--(2biA)--(SCS);
        \draw[-stealth, thick](2bi)--(2biB)--(CYCY);
       \draw[-stealth, thick](2b)--(2bii);
        \draw[-stealth, thick](2bii)--(2biiA)--(BHM1-3);
        \draw[-stealth, thick](2bii)--(2biiB);
         \draw[-stealth, thick](2biiB)--(I);
         \draw[-stealth, thick](2biiB)--(II);
      \draw[-stealth, thick](2)--(2c);
      \draw[-stealth, thick](2c)--(fd);
      \draw[-stealth, thick](2)--(2d) to 
        node[above, xshift=-1mm, rotate=-88]
        {\scriptsize\it foliation}(fo);
     \draw[-stealth, densely dashed, thick](S)--(3)--(3e)--(5) to 
        node[above, rotate=-90]{\scriptsize\it Lorentzian}(HH);
     \path(.45,-5.4) node {\scriptsize rigid product};
     \path(2.55,-5.4) node {\scriptsize stringy cosmic branes};
     \path(5.5,-5.4) node {\footnotesize\cite{Berglund:2001wj,Berglund:2001aj,Berglund:2019pxr,Berglund:2020qcu}};
     \node(HHr) at(10,-5.4)
        {\footnotesize\cite{Hubsch:1997ty,Berglund:2021xlm}};
     \
     \draw[densely dashed,-stealth,thick]
             (1,-5.6)to[out=-45,in=-135](1.8,-5.6);
     \draw[densely dashed,-stealth,thick]
             (3.2,-5.6)to[out=-45,in=-135](4,-5.35)to[out=-45,in=-165](4.8,-5.5);
     \draw[densely dashed,stealth-stealth,thick]
             (6.25,-5.4)to[out=-5,in=-180](A-dS);
     \draw[densely dashed,stealth-stealth,thick]
             (A-dS)to[out=0,in=-175](HHr);
   \end{tikzpicture}
 \caption{A (partial) roadmap with some spacetime-variable compactification constructions in the literature~\cite{Berglund:2022qsb}, lined up from the simplest at the far left to the more and more complex (and speculative) at the far right, and with a hierarchy of overarching relations tracing their lineage}
 \label{f:FTree}
\end{figure}
\paragraph{And More, Much More:}
Figure~\ref{f:FTree} presents a palette of possibilities for stringy spacetime discussed (if tentatively) in the literature, where the compactification space varies over the directly observable spacetime, $M^{1,3}$ in increasingly more complex ways. 
Towards the far right in Figure~\ref{f:FTree}, the oft-presumed notion that the compact space $X$, as much as it may vary\footnote{\label{fn:sstrs}What varies are its string-relevant: complex, Hodge, complexified K\"ahler, Hull-Strominger, HYM, special Lagrangian, etc.\ structures, and the induced phenomenology-relevant: seesaw, Froggatt--Nielsen, and other details in the physically normalized mass-hierarchy structure of the Standard Model~\cite{Berglund:2023gur,Hubsch:2024agh}.} over $M^{1,3}$, remains too small to be directly observable becomes untenable: in a {\em\/foliation,} even the product structure is defined only locally. The ultimate, rightmost-depicted option, $\mathfrak{X}^{1,9}$, has no such product structure; instead, the observable spacetime is here modeled as a (dynamically~\cite{Banerjee:2018qey,Blaback:2019zig} and/or quasi-topologically~\cite{Hubsch:1997ty,Berglund:2021xlm}) localized subspace (together with {\em us\/}: the matter and gauge interactions of the Standard Model) {\em within\/} a much bigger Ricci-flat spacetime, $M^{1,3}\subset\mathfrak{X}^{1,9}$.

These increasingly more speculative proposals however still build upon the more concrete ``stringy cosmic braneworld'' scenarios~\cite{Randall:1999vf,Berglund:2021xlm}, and so could be made much more concrete by numerical solutions to the metric and other stringy structures; see footnote~\ref{fn:sstrs}.
This of course requires solving Lorentzian field equations --- which is anyway the ultimate goal of modeling the Universe at its most fundamental level, to which end a suitably Wick-rotated Riemannian Calabi--Yau metric could serve as a starting point. 
The oft-cited topological requirement, that a compact smooth manifold admits a {\em\/nonsingular\/} Lorentzian metric when it admits a nowhere-vanishing vector field, which is on a compact orientable manifolds tied to its Euler characteristic vanishing~\cite{Geroch:1967fs} --- is a red herring: 
Many physically interesting classical (non-quantum) cosmological solutions do exhibit (both in the underlying space and the time-like vector field on it) at least one singularity at the ``beginning of time,'' and (if the Universe continues to expand accelerating) may well have multiple singularities at ``infinite future time,'' neither of which need be points. Thus, the time-flow vector field {\em may well be singular,} at the ``infinite past'' and the ``infinite future,'' lest one seeks a cyclic time-flow.  Numerical Calabi--Yau metrics can nevertheless provide controlled internal data for developing and testing such Ans\"atze.

\subsubsection{Open problems}

\begin{enumerate}[label=(\roman*)]
\item \textbf{Numerical spacetime metrics from varying Calabi--Yau moduli:}
Given a map from the observable spacetime into the moduli space of a compactification, compute the induced Weil--Petersson--Zamolodchikov metric, warping, and Ricci curvature, and determine when the resulting spacetime solves the higher-dimensional field equations to controlled accuracy.

\item \textbf{Metric behavior near discriminant loci:}
Analyze the numerical metric near the real codimension-two loci where the varying \CY fiber singularizes, and compute the associated monodromy, curvature concentration, deficit angle, effective tension, and possible matter accretion profile of the resulting stringy cosmic strings.

\item \textbf{Compact/non-compact matching in the ambient geometry:}
Compute and compare the numerical metrics $\num{g}_X$, $\num{g}_A$, and $\num{g}_{A^{nc}}$ in the compact fiber, ambient Fano space, and non-compact \CY complement, and test the asymptotic relations among $J_X$, $J_A$, $J_{A^{nc}}$, $\Omega_X$, and $\Omega_{A^{nc}}$ near the fiber at infinity.

\item \textbf{Observable signatures of stringy cosmic branes:}
Determine which features of the induced spacetime geometry, such as lensing, gravitational backreaction, accretion of visible or dark matter, or filamentary large-scale structure, are robust consequences of spacetime variable compactification rather than adjustable phenomenological assumptions.

\item \textbf{Lorentzian Ans\"atze from higher-dimensional internal data:}
Develop criteria for when numerical Calabi--Yau data on higher-dimensional spaces can be used as initial or internal data for Lorentzian equations, rather than assuming that a Riemannian Ricci-flat metric can simply be Wick-rotated.

\item \textbf{Localized de~Sitter-like regions:}
Determine whether candidate localized four-dimensional regions associated with special subvarieties or singular sources can support a controlled Lorentzian de~Sitter-like metric after including embedding effects, extrinsic curvature, higher-dimensional backreaction, and possible source terms.

\item \textbf{Height functions and causal structure:}
For compact or non-compact spaces inspired by Calabi--Yau geometry, identify whether suitable height functions or direct Lorentzian numerical Ansatz can satisfy the field equations and causal constraints, including the exclusion of closed timelike curves and the treatment of boundaries or singular sources.

\item \textbf{Stability and localization of spacetime bubbles:}
Analyze the perturbative stability, matter localization, and backreaction of candidate spacetime bubbles, and determine whether the internal metric data can produce a four-dimensional effective theory with a sensible hierarchy between localized spacetime dynamics and the surrounding higher-dimensional geometry.
\end{enumerate}

\section{Conclusions and outlook}
\label{s:CODA}
The long-term significance of computable Ricci-flat \CY metrics is not merely that they allow technical progress.
It is that they change the character of the subject.
String phenomenology has often been forced to operate with a partial view of the compactification: topology, intersection theory, cohomology, periods, holomorphy, and protected quantities were accessible, while the metric dependent data that determine canonical normalization, physical scales, localization, and the validity of approximations remained largely implicit.
A precise approximation to the Ricci-flat metric, with quantified error bars, turns the compactification from a formal background into a bona fide {\em geometric\/} object.
The future of the field should therefore not just be organized only around finding more vacua, but around asking which vacua can be made computationally sharp enough that their effective field theories are determined with meaningful uncertainties.

This shift suggests a more stringent standard for phenomenological model building.
A candidate compactification should eventually come with a numerical package: the metric, the relevant bundle or brane data, harmonic representatives, normalized couplings, Kaluza--Klein spectra, curvature diagnostics, and moduli dependence, all accompanied by uncertainties.
Such a package would make it possible to distinguish qualitative plausibility from quantitative prediction.
Hierarchies in Yukawa couplings, axion decay constants, soft terms, threshold effects, and towers of states would no longer be implicit; they could be measured inside the compact space.
Conversely, many attractive scenarios may fail precisely because the metric reveals that a presumed hierarchy is not present or has the phenomenologically unacceptable scale, that curvature is too large or too small, that a localization argument is too crude, or that a controlled effective description breaks down.
This is a feature rather than a defect: Numerical Ricci-flat geometry will 
make string phenomenology more falsifiable.
Pushing this program ahead, we can conceive that in the near future, we will have examples of string Standard Models consistent with cosmology in which hierarchies are stabilized.
In our view, it is likely that if one such example exists, there will be many.
The vacuum selection problem does not go away, but working examples of ``real world''-like theories may allow us to test correlations among parameters in a statistically meaningful way and define the borders of the string landscape~\cite{Denef:2004ze,Denef:2006ad}.

There is an equally important mathematical consequence.
Yau's theorem proves existence and uniqueness, but a computable metric gives experimental/computational access to the object whose existence is guaranteed.
This creates a new kind of {\em metric \CY geometry\/} in which one can formulate conjectures from data: about the distribution of curvature, the behavior of Laplace spectra, the geometry of degenerating cycles, the emergence and geometric details of special Lagrangian fibrations, the structure of high curvature regions, and the variation of these quantities across moduli space.
The role of computation here is not to replace proof.
Rather, it is to expose regularities that were invisible from purely algebraic data, to test the sharpness of asymptotic theorems, and to suggest which analytic approximations are worth trying to prove.
In favorable cases, machine learned or balanced metric representations may even be compressed into semi-analytic formul\ae, turning numerical discovery into explicit geometric structure.

The most exciting future direction is the convergence of these viewpoints.
The same metric data needed for physical normalization also probes deep mathematical questions about mirror symmetry, SYZ fibrations, calibrated submanifolds, and metric degeneration.
The same moduli dependent family needed for phenomenology also supplies a laboratory for geodesic distance, monodromy, collapsing limits, and emergent light states.
The same error diagnostics that make numerical predictions trustworthy also define benchmarks for geometric approximation schemes.
In this sense, Ricci-flat metrics are not just another ingredient in the compactification pipeline; they are the common language in which phenomenological precision and metric geometry interact.

The guiding lesson is simple: once the Ricci-flat metric is in hand, the \CY is no longer a space whose existence is understood only abstractly.
It becomes a precise geometry.
The next stage of string phenomenology and \CY mathematics will be shaped by learning how to use this precept in an engineering sense.

\paragraph{Acknowledgments:}
We thank the Chennai Mathematical Institute for hosting the BIRS workshop ``Recent Progress in Computational String Geometry'' in January 2026~\cite{BIRS:2026}, our co-organizers Dami\'an Mayorga Pe\~na, Challenger Mishra, and Ivonne Zavala, and the participants at the meeting.
We are grateful for collaborations and conversations with Giorgi Butbaia, Dami\'an Mayorga Pe\~na, Viktor Mirjani\'c, Challenger Mishra, Justin Tan, and Simon Tollman.
We acknowledge the use of GPT-5.4 and GPT-5.5 in preparing parts of this manuscript.
The authors take full responsibility for the accuracy of this work, having reviewed, verified, and approved all AI-generated content.
PB thanks the CERN Theory Group for their hospitality and the support of the U.S.\ Department of Energy grant DE-SC0020220.
TH is grateful to the Department of Mathematics, University of Maryland, College Park, and the Physics Department of the Faculty of Natural Sciences of the University of Novi Sad, Serbia, for the recurring hospitality and resources.
VJ is supported by the South African Research Chairs Initiative of the Department of Science, Technology, and Innovation and the National Research Foundation (grant 78554).

\bibliographystyle{JHEP}
\bibliography{refs.bib}

@article{Berglund:2023gur,
	archiveprefix = {arXiv},
	author = {Berglund, Per and H{\"u}bsch, Tristan and Minic, Djordje},
	doi = {10.3390/sym15091660},
	eprint = {2307.16712},
	journal = {Symmetry},
	month = {Aug.},
	number = {9},
	pages = {1660},
	primaryclass = {hep-th},
	title = {String Theory Bounds on the Cosmological Constant, the {H}iggs Mass, and the Quark and Lepton Masses},
	volume = {15},
	year = {2023}}

@inproceedings{Hubsch:2024agh,
	address = {Belgrade, Serbia},
	archiveprefix = {arXiv},
	author = {H{\"u}bsch, Tristan and Minic, Djordje},
	booktitle = {Proceedings of the 3rd Conference on Nolinearity (Sep. 4--8.\ 2023)},
	editor = {B. Dragovich and {\v{Z}}. {\v{C}}upi{\'c}},
	eprint = {2407.06207},
	pages = {1--72},
	primaryclass = {hep-th},
	publisher = {Serbian Academy of Nonlinear Sciences},
	title = {Quantum Gravity as Gravitized Quantum Theory},
	year = {2024}}

@article{Berglund:2022qsb,
    author = {Berglund, Per and H{\"u}bsch, Tristan and Minic, Djordje},
    title = "{On de Sitter Spacetime and String Theory}",
    eprint = "2212.06086",
    archivePrefix = "arXiv",
    primaryClass = "hep-th",
    doi = "10.1142/S0218271823300021",
    journal = "Int. J. Mod. Phys. D",
    volume = "32",
    pages = "2330002",
    year = "2023"
}

@article{Banerjee:2018qey,
	archiveprefix = {arXiv},
	author = {Banerjee, Souvik and Danielsson, Ulf and Dibitetto, Giuseppe and Giri, Suvendu and Schillo, Marjorie},
	doi = {10.1103/physrevlett.121.261301},
	eprint = {1807.01570},
	journal = {Phys. Rev. Lett.},
	month = {Dec},
	number = {26},
	pages = {261301 (6p)},
	primaryclass = {hep-th},
	title = {Emergent de~{S}itter Cosmology from Decaying Anti--de~{S}itter Space},
	volume = {121},
	year = {2018}}

@article{Blaback:2019zig,
	archiveprefix = {arXiv},
	author = {Bl{\aa}b{\"a}ck, Johan and Danielsson, Ulf and Dibitetto, Giuseppe and Giri, Suvendu},
	eprint = {1902.04053},
	primaryclass = {hep-th},
	title = {Constructing stable de~{S}itter in {M}-theory from higher curvature corrections},
	year = {2019}}

@article{Berglund:2020qcu,
	archiveprefix = {arXiv},
	author = {Berglund, Per and H{\"u}bsch, Tristan and Minic, Djordje},
	doi = {10.31526/lhep.2021.186},
	eprint = {2010.15610},
	journal = {LHEP},
	pages = {186},
	primaryclass = {hep-th},
	title = {String Theory, the Dark Sector and the Hierarchy Problem},
	volume = {2021},
	year = {2021}}

@article{Berglund:2000gt,
	archiveprefix = {arXiv},
	author = {Berglund, Per and Hubsch, Tristan and Minic, D.},
	doi = {10.1088/1126-6708/2001/02/010},
	eprint = {hep-th/0012042},
	journal = {JHEP},
	pages = {010},
	title = {Probing naked singularities in nonsupersymmetric string vacua},
	volume = {02},
	year = {2001}}

@article{Berglund:2000mr,
	archiveprefix = {arXiv},
	author = {Berglund, Per and Hubsch, Tristan and Minic, D.},
	doi = {10.1088/1126-6708/2001/01/041},
	eprint = {hep-th/0012180},
	journal = {JHEP},
	pages = {041},
	title = {On Relativistic brane probes in singular space-times},
	volume = {01},
	year = {2001}}

@article{Berglund:2001wj,
	archiveprefix = {arXiv},
	author = {Berglund, Per and Hubsch, Tristan and Minic, D.},
	doi = {10.1016/S0370-2693(01)00665-7},
	eprint = {hep-th/0104057},
	journal = {Phys. Lett. B},
	pages = {155--160},
	title = {Localized gravity and large hierarchy from string theory?},
	volume = {512},
	year = {2001}}

@article{MacFadden:2025ssx,
	archiveprefix = {arXiv},
	author = {MacFadden, Nate and Sheridan, Elijah},
	eprint = {2512.14817},
	month = {12},
	primaryclass = {hep-th},
	title = {{C}alabi--{Y}au Threefolds from Vex Triangulations},
	year = {2025}}

@article{MacFadden:2024him,
	archiveprefix = {arXiv},
	author = {MacFadden, Nate and Schachner, Andreas and Sheridan, Elijah},
	eprint = {2405.08871},
	month = {5},
	primaryclass = {hep-th},
	title = {The {DNA} of {C}alabi--{Y}au Hypersurfaces},
	year = {2024}}

@article{Avram:1997rs,
	archiveprefix = {arXiv},
	author = {Avram, A. C. and Kreuzer, M. and Mandelberg, M. and Skarke, H.},
	doi = {10.1016/S0550-3213(97)00582-8},
	eprint = {hep-th/9703003},
	journal = {Nucl. Phys.},
	pages = {625-640},
	primaryclass = {hep-th},
	title = {The Web of {C}alabi--{Y}au hypersurfaces in toric varieties},
	volume = {B505},
	year = {1997}}

@article{Avram:1995pu,
	archiveprefix = {arXiv},
	author = {Avram, A. C. and Candelas, P. and Jancic, D. and Mandelberg, M.},
	doi = {10.1016/0550-3213(96)00058-2},
	eprint = {hep-th/9511230},
	journal = {Nucl. Phys.},
	pages = {458-472},
	primaryclass = {hep-th},
	title = {On the connectedness of moduli spaces of {C}alabi--{Y}au manifolds},
	volume = {B465},
	year = {1996}}

@article{Candelas:1989qn,
	author = {Candelas, Philip and H{\"u}bsch, Tristan and Schimmrigk, Rolf},
	doi = {10.1016/0550-3213(90)90072-L},
	journal = {Nucl. Phys. B},
	pages = {583--590},
	title = {Relation Between the {W}eil--{P}etersson and {Z}amolodchikov Metrics},
	volume = {329},
	year = {1990}}

@article{Green:1988bp,
    author = "Green, Paul S. and H{\"u}bsch, Tristan",
    title = "{Connecting Moduli Spaces of Calabi-yau Threefolds}",
    reportNumber = "UTTG-4-88",
    doi = "10.1007/BF01218081",
    journal = "Commun. Math. Phys.",
    volume = "119",
    pages = "431--441",
    year = "1988"
}

@article{Berglund:2022dgb,
	archiveprefix = {arXiv},
	author = {Berglund, Per and H{\"u}bsch, Tristan},
	doi = {10.4310/ATMP.2022.v26.n8.a3},
	eprint = {2205.12827},
	journal = {Adv. Theor. Math. Phys.},
	number = {8},
	pages = {2541-2598},
	primaryclass = {hep-th},
	title = {Hirzebruch Surfaces, {T}yurin Degenerations and Toric Mirrors: Bridging Generalized {C}alabi--{Y}au Constructions},
	volume = {26},
	year = {2022}}

@inproceedings{tyurin2003fano,
	address = {Univ. Torino, Turin},
	archiveprefix = {arXiv},
	author = {Andrey N. Tyurin},
	booktitle = {The Fano Conference},
	eprint = {math/0302101},
	pages = {701--734},
	primaryclass = {math.AG},
	title = {Fano versus {C}alabi--{Y}au},
	year = {2004}}

@article{Tian:1991swq,
    author = {Tian, Gang and Yau, Shing Tung},
    title = {Complete {K}{\"a}hler manifolds with zero Ricci curvature II},
    doi = {10.1007/BF01243902},
    journal = {Invent. Math.},
    volume = {106},
    number = {1},
    pages = {27--60},
    year = {1991}
}

@article{Tian:1990ty,
    author = {Tian, Gang and Yau, Shing-Tung},
    title = {Complete {K}{\"a}hler manifolds with zero {R}icci curvature. {I}.},
    doi = {10.2307/1990928},
    journal = {J. Amer. Math. Soc.},
    volume = {3},
    number = {3},
    pages = {579--609},
    year = {1990}
}

@article{Candelas:1989ug,
    author = {Candelas, Philip and Green, Paul S. and H{\"u}bsch, Tristan},
    title = {Rolling Among {C}alabi--{Y}au Vacua},
    reportNumber = {UTTG-10-89},
    doi = {10.1016/0550-3213(90)90302-T},
    journal = {Nucl. Phys. B},
    volume = {330},
    pages = {49},
    year = {1990}
}

@article{Green:1988wa,
    author = {Green, Paul S. and H{\"u}bsch, Tristan},
    title = {Possible Phase Transitions among {C}alabi--{Y}au Compactifications},
    reportNumber = {UTTG-06-88},
    doi = {10.1103/PhysRevLett.61.1163},
    journal = {Phys. Rev. Lett.},
    volume = {61},
    pages = {1163},
    year = {1988}
}

@article{Anderson:2022bpo,
    author = {Anderson, Lara B. and Brodie, Callum R. and Gray, James},
    title = {Branes and Bundles through Conifold Transitions and Dualities in Heterotic String Theory},
    eprint = {2211.05804},
    archivePrefix = {arXiv},
    primaryClass = {hep-th},
    journal = {Phys. Rev. D},
    volume = {108},
    number = {10},
    pages = {106019},
    year = {2023}
}

@article{Berglund:2019pxr,
    author = {Berglund, Per and H{\"u}bsch, Tristan and Minic, Djordje},
    title = {On Stringy de~{S}itter Spacetimes},
    eprint = {1902.08617},
    archivePrefix = {arXiv},
    primaryClass = {hep-th},
    doi = {10.1007/JHEP12(2019)166},
    journal = {JHEP},
    volume = {12},
    pages = {166},
    year = {2019}
}

@article{Berglund:2001aj,
    author = {Berglund, Per and H{\"u}bsch, Tristan and Minic, D.},
    title = {De~{S}itter space-times from warped compactifications of {IIB} string theory},
    eprint = {hep-th/0112079},
    archivePrefix = {arXiv},
    reportNumber = {CUTUSC-01-47, VPI-IPPAP-01-05},
    doi = {10.1016/S0370-2693(02)01713-6},
    journal = {Phys. Lett. B},
    volume = {534},
    pages = {147--154},
    year = {2002}
}

@article{Berglund:2000fy,
    author = {Berglund, Per and H{\"u}bsch, Tristan and Minic, D.},
    title = {Exponential hierarchy from space-time variable string vacua},
    eprint = {hep-th/0005162},
    archivePrefix = {arXiv},
    reportNumber = {NSF-ITP-00-37, CITUSC-00-022},
    doi = {10.1088/1126-6708/2000/09/015},
    journal = {JHEP},
    volume = {09},
    pages = {015},
    year = {2000}
}

@article{Berglund:2021xlm,
    author = {Berglund, Per and H{\"u}bsch, Tristan and Minic, Djordje},
    title = {Stringy Bubbles Solve de~{S}itter Troubles},
    eprint = {2109.01122},
    archivePrefix = {arXiv},
    primaryClass = {hep-th},
    doi = {10.3390/universe7100363},
    journal = {Universe},
    volume = {7},
    number = {10},
    pages = {363},
    year = {2021}
}

@article{Hubsch:1997ty,
    author = {H{\"u}bsch, Tristan},
    editor = {Mohapatra, R. N. and Rasin, A.},
    title = {A hitchhiker's guide to superstring jump gates and other worlds},
    doi = {10.1016/S0920-5632(96)00589-0},
    journal = {Nucl. Phys. B Proc. Suppl.},
    volume = {52},
    pages = {347--350},
    year = {1997}
}

@article{Randall:1999vf,
    author = {Randall, Lisa and Sundrum, Raman},
    title = {An Alternative to compactification},
    eprint = {hep-th/9906064},
    archivePrefix = {arXiv},
    doi = {10.1103/PhysRevLett.83.4690},
    journal = {Phys. Rev. Lett.},
    volume = {83},
    pages = {4690--4693},
    year = {1999}
}

@article{Kreuzer:2000xy,
    author = {Kreuzer, Maximilian and Skarke, Harald},
    title = {Complete classification of reflexive polyhedra in four-dimensions},
    eprint = {hep-th/0002240},
    archivePrefix = {arXiv},
    doi = {10.4310/ATMP.2000.v4.n6.a2},
    journal = {Adv. Theor. Math. Phys.},
    volume = {4},
    pages = {1209--1230},
    year = {2000}
}

@article{Green:1993zr,
    author = {Green, Paul S. and H{\"u}bsch, Tristan},
    title = {Space-time variable superstring vacua ({C}alabi--{Y}au cosmic yarn)},
    eprint = {hep-th/9306057},
    archivePrefix = {arXiv},
    reportNumber = {HUPAPP-93-1},
    doi = {10.1142/S0217751X94001266},
    journal = {Int. J. Mod. Phys. A},
    volume = {9},
    pages = {3203--3228},
    year = {1994}
}

@article{Greene:1989ya,
    author = {Greene, Brian R. and Shapere, Alfred D. and Cumrun Vafa and Yau, Shing-Tung},
    title = {Stringy Cosmic Strings and Noncompact {C}alabi--{Y}au Manifolds},
    doi = {10.1016/0550-3213(90)90248-C},
    journal = {Nucl. Phys. B},
    volume = {337},
    pages = {1--36},
    year = {1990}
}

@article{Yau:1978rc,
    author = {Yau, Shing-Tung},
    title = {On the {R}icci Curvature of a Compact {K}{\"a}hler Manifold and the Complex {M}onge--{A}mp{\`e}re Equation. {I}},
    doi = {10.1002/cpa.3160310304},
    journal = {Commun. Pure Appl. Math.},
    volume = {31},
    number = {3},
    pages = {339--411},
    year = {1978}
}

@article{Anderson:2010ke,
    author = {Anderson, Lara B. and Braun, Volker and Karp, Robert L. and Ovrut, Burt A.},
    title = {Numerical Hermitian {Y}ang--{M}ills Connections and Vector Bundle Stability in Heterotic Theories},
    eprint = {1004.4399},
    archivePrefix = {arXiv},
    primaryClass = {hep-th},
    doi = {10.1007/JHEP06(2010)107},
    journal = {JHEP},
    volume = {06},
    pages = {107},
    year = {2010}
}

@article{Becker:2005nb,
    author = {Becker, Katrin and Tseng, Li-Sheng},
    title = {Heterotic Flux Compactifications and Their Moduli},
    eprint = {hep-th/0509131},
    archivePrefix = {arXiv},
    primaryClass = {hep-th},
    doi = {10.1016/j.nuclphysb.2006.02.013},
    journal = {Nucl. Phys. B},
    volume = {741},
    number = {1--2},
    pages = {162--179},
    year = {2006}
}

@article{Candelas:2016usb,
    author = {Candelas, Philip and de la Ossa, Xenia and McOrist, Jock},
    title = {A Metric for Heterotic Moduli},
    eprint = {1605.05256},
    archivePrefix = {arXiv},
    primaryClass = {hep-th},
    doi = {10.1007/s00220-017-2978-7},
    journal = {Commun. Math. Phys.},
    volume = {356},
    number = {2},
    pages = {567--612},
    year = {2017}
}

@article{Becker:2002nn,
    author = {Becker, Katrin and Becker, Melanie and Haack, Michael and Louis, Jan},
    title = {Supersymmetry Breaking and alpha$'$-Corrections to Flux Induced Potentials},
    eprint = {hep-th/0204254},
    archivePrefix = {arXiv},
    primaryClass = {hep-th},
    doi = {10.1088/1126-6708/2002/06/060},
    journal = {JHEP},
    volume = {06},
    pages = {060},
    year = {2002}
}

@article{McAllister:2024lnt,
    author = {McAllister, Liam and Moritz, Jakob and Nally, Richard and Schachner, Andreas},
    title = {Candidate de~{S}itter Vacua},
    eprint = {2406.13751},
    archivePrefix = {arXiv},
    primaryClass = {hep-th},
    doi = {10.1103/PhysRevD.111.086015},
    journal = {Phys. Rev. D},
    volume = {111},
    number = {8},
    pages = {086015},
    year = {2025}
}

@article{Dine:2020vmr,
    author = {Dine, Michael and Law-Smith, Jamie A. P. and Sun, Shijun and Wood, Duncan and Yu, Yan},
    title = {Obstacles to Constructing de~{S}itter Space in String Theory},
    eprint = {2008.12399},
    archivePrefix = {arXiv},
    primaryClass = {hep-th},
    doi = {10.1007/JHEP02(2021)050},
    journal = {JHEP},
    volume = {02},
    pages = {050},
    year = {2021}
}

@article{Candelas:1989js,
    author = {Candelas, Philip and de la Ossa, Xenia C.},
    title = {Comments on conifolds},
    doi = {10.1016/0550-3213(90)90577-Z},
    journal = {Nucl. Phys. B},
    volume = {342},
    number = {1},
    pages = {246--268},
    year = {1990}
}

@article{Stenzel:1993rf,
    author = {Stenzel, Matthew B.},
    title = {Ricci-flat metrics on the complexification of a compact rank one symmetric space},
    doi = {10.1007/BF03026543},
    journal = {Manuscripta Math.},
    volume = {80},
    number = {1},
    pages = {151--163},
    year = {1993}
}

@article{Gross:2000wp,
    author = {Gross, Mark and Wilson, P. M. H.},
    title = {Large complex structure limits of {K3} surfaces},
    eprint = {math/0008018},
    archivePrefix = {arXiv},
    doi = {10.4310/jdg/1090341262},
    journal = {J. Diff. Geom.},
    volume = {55},
    number = {3},
    pages = {475--546},
    year = {2000}
}

@article{Strominger:1996it,
    author = {Strominger, Andrew and Yau, Shing-Tung and Zaslow, Eric},
    title = {Mirror Symmetry is {T}-Duality},
    eprint = {hep-th/9606040},
    archivePrefix = {arXiv},
    primaryClass = {hep-th},
    doi = {10.1016/0550-3213(96)00434-8},
    journal = {Nucl. Phys. B},
    volume = {479},
    pages = {243--259},
    year = {1996}
}

@article{Joyce:2000js,
    author = {Joyce, Dominic},
    title = {Singularities of Special Lagrangian Fibrations and the {SYZ} Conjecture},
    eprint = {math/0011179},
    archivePrefix = {arXiv},
    primaryClass = {math.DG},
    journal = {Commun. Anal. Geom.},
    volume = {11},
    number = {5},
    pages = {859--907},
    year = {2003}
}

@article{Gross:1999tm,
    author = {Gross, Mark},
    title = {Topological Mirror Symmetry},
    eprint = {math/9909015},
    archivePrefix = {arXiv},
    primaryClass = {math.AG},
    doi = {10.1007/s002220000119},
    journal = {Invent. Math.},
    volume = {144},
    number = {1},
    pages = {75--137},
    year = {2001}
}

@article{Demirtas:2023als,
    author = {Demirtas, Mehmet and Kim, Manki and McAllister, Liam and Moritz, Jakob and Rios-Tascon, Andres},
    title = {Computational Mirror Symmetry},
    eprint = {2303.00757},
    archivePrefix = {arXiv},
    primaryClass = {hep-th},
    reportNumber = {MIT-CTP/5528},
    doi = {10.1007/JHEP01(2024)184},
    journal = {JHEP},
    volume = {01},
    pages = {184},
    year = {2024}
}

@article{Jejjala:2020wcc,
    author = {Jejjala, Vishnu and Mayorga Pe{\~n}a, Dami{\'a}n Kaloni and Mishra, Challenger},
    title = {Neural network approximations for {C}alabi--{Y}au metrics},
    eprint = {2012.15821},
    archivePrefix = {arXiv},
    primaryClass = {hep-th},
    doi = {10.1007/JHEP08(2022)105},
    journal = {JHEP},
    volume = {08},
    pages = {105},
    year = {2022}
}

@inproceedings{Douglas:2020hpv,
	archiveprefix = {arXiv},
	author = {Douglas, Michael R. and Lakshminarasimhan, Subramanian and Qi, Yidi},
    booktitle={Proceedings of the 2nd Mathematical and Scientific Machine Learning Conference},
    editor={Bruna, Joan and Hesthaven, Jan and Zdeborova, Lenka},
	eprint = {2012.04797},
	journal = {Proc. ML Res.},
	note = {2nd Annual Conference on Mathematical and Scientific Machine Learning, Joan Bruna and Jan S Hesthaven and Lenka Zdeborova, eds.},
	pages = {223--252},
	primaryclass = {hep-th},
	title = {Numerical {C}alabi--{Y}au metrics from holomorphic networks},
	volume = {145},
	year = {2022}}

@article{Douglas:2021zdn,
    author = {Douglas, Michael R.},
    title = {Holomorphic feedforward networks},
    eprint = {2105.03991},
    archivePrefix = {arXiv},
    primaryClass = {math.CV},
    doi = {10.4310/PAMQ.2022.v18.n1.a7},
    journal = {Pure Appl. Math. Quart.},
    volume = {18},
    number = {1},
    pages = {251--268},
    year = {2022}
}

@article{Headrick:2005ch,
    author = {Headrick, Matthew and Wiseman, Toby},
    title = {Numerical {R}icci-flat metrics on {K3}},
    eprint = {hep-th/0506129},
    archivePrefix = {arXiv},
    reportNumber = {HUTP-05-A0028, MIT-CTP-3647},
    doi = {10.1088/0264-9381/22/23/002},
    journal = {Class. Quant. Grav.},
    volume = {22},
    pages = {4931--4960},
    year = {2005}
}

@article{Donaldson:2005hvr,
	archiveprefix = {arXiv},
	author = {Donaldson, S. K.},
	doi = {10.4310/PAMQ.2009.v5.n2.a2},
	eprint = {math/0512625},
	journal = {Pure Appl. Math. Quart.},
	month = {12},
	number = {2},
	pages = {571-618},
	title = {Some numerical results in complex differential geometry},
	volume = {5},
	year = {2009}}

@article{Gerdes:2022nzr,
    author = {Gerdes, Mathis and Krippendorf, Sven},
    title = {{CYJAX}: A package for {C}alabi--{Y}au metrics with {JAX}},
    eprint = {2211.12520},
    archivePrefix = {arXiv},
    primaryClass = {hep-th},
    doi = {10.1088/2632-2153/acdc84},
    journal = {Mach. Learn. Sci. Tech.},
    volume = {4},
    number = {2},
    pages = {025031},
    year = {2023}
}

@book{Hubsch:2024cym,
    author = {Tristan H{\"u}bsch},
    title = {{C}alabi--{Y}au Manifolds: a Bestiary for Physicists},
    year = {2024 (1st ed., 1992, {W}orld {S}cientific {P}ub., Singapore)},
    publisher = {World Scientific Pub. Europe Ltd.},
    address = {London, UK},
    edition = {2nd}
}

@article{Berglund:2023ztk,
    author = {Berglund, Per and He, Yang-Hui and Heyes, Elli and Hirst, Edward and Jejjala, Vishnu and Lukas, Andre},
    title = {New {C}alabi--{Y}au Manifolds from Genetic Algorithms},
    eprint = {2306.06159},
    archivePrefix = {arXiv},
    primaryClass = {hep-th},
    doi = {10.1016/j.physletb.2024.138504},
    journal = {Phys. Lett. B},
    volume = {850},
    pages = {138504},
    year = {2024}
}

@article{Berglund:2022gvm,
    author = {Berglund, Per and Butbaia, Giorgi and H{\"u}bsch, Tristan and Jejjala, Vishnu and Mayorga Pe{\~n}a, Dami{\'a}n and Mishra, Challenger and Tan, Justin},
    title = {Machine Learned {C}alabi--{Y}au Metrics and Curvature},
    eprint = {2211.09801},
    archivePrefix = {arXiv},
    primaryClass = {hep-th},
    doi = {10.4310/ATMP.2023.v27.n4.a3},
    journal = {Adv. Theor. Math. Phys.},
    volume = {27},
    number = {4},
    pages = {1107-1158},
    year = {2023}
}

@article{Berglund:2024psp,
    author = {Butbaia, Giorgi and Mayorga Pe{\~n}a, Dami{\'a}n and Tan, Justin and Berglund, Per and H{\"u}bsch, Tristan and Jejjala, Vishnu and Mishra, Challenger},
    title = {cymyc: {C}alabi--{Y}au Metrics, {Y}ukawas, and Curvature},
    eprint = {2410.19728},
    archivePrefix = {arXiv},
    primaryClass = {hep-th},
    doi = {10.1007/JHEP03(2025)028},
    journal = {JHEP},
    volume = {03},
    pages = {28},
    year = {2025}
}

@article{Berglund:2024reu,
    author = {Berglund, Per and Butbaia, Giorgi and He, Yang-Hui and Heyes, Elli and Hirst, Edward and Jejjala, Vishnu},
    title = {Generating triangulations and fibrations with reinforcement learning},
    eprint = {2405.21017},
    archivePrefix = {arXiv},
    primaryClass = {hep-th},
    reportNumber = {QMUL-PH-24-10},
    doi = {10.1016/j.physletb.2024.139158},
    journal = {Phys. Lett. B},
    volume = {860},
    pages = {139158},
    year = {2025}
}

@inproceedings{Larfors:2021pbb,
	archiveprefix = {arXiv},
	author = {Larfors, Magdalena and Lukas, Andre and Ruehle, Fabian and Schneider, Robin},
	title = {Learning Size and Shape of {C}alabi--{Y}au Spaces},
	booktitle = {Machine Learning and the Physical Sciences, Workshop at 35th {NeurIPS}},
    eprint = {2111.01436},
    archivePrefix = {arXiv},
    primaryClass = {hep-th},
    reportNumber = {UUITP-53/21},
    year = {2021},
    month = {11}
}

@article{Headrick:2009jz,
    author = {Headrick, Matthew and Nassar, Ali},
    title = {Energy functionals for {C}alabi--{Y}au metrics},
    eprint = {0908.2635},
    archivePrefix = {arXiv},
    primaryClass = {hep-th},
    reportNumber = {BRX-TH-612},
    doi = {10.4310/ATMP.2013.v17.n5.a1},
    journal = {Adv. Theor. Math. Phys.},
    volume = {17},
    number = {5},
    pages = {867--902},
    year = {2013}
}

@article{Yau:1977ms,
    author = {Yau, Shing-Tung},
    title = {Calabi's conjecture and some new results in algebraic geometry},
    doi = {10.1073/pnas.74.5.1798},
    journal = {Proc. Nat. Acad. Sci.},
    volume = {74},
    pages = {1798--1799},
    year = {1977}
}

@article{Butbaia:2024tje,
    author = {Butbaia, Giorgi and Mayorga Pe\~na, Dami\'an and Tan, Justin and Berglund, Per and H\"ubsch, Tristan and Jejjala, Vishnu and Mishra, Challenger},
    title = {Physical {Y}ukawa Couplings in Heterotic String Compactifications},
    eprint = {2401.15078},
    archivePrefix = {arXiv},
    primaryClass = {hep-th},
    doi = {10.4310/ATMP.241119041341},
    journal = {Adv. Theor. Math. Phys.},
    volume = {28},
    number = {8},
    pages = {2783--2822},
    year = {2024}
}

@article{Berglund:2024uqv,
    author = {Berglund, Per and Butbaia, Giorgi and H\"ubsch, Tristan and Jejjala, Vishnu and Mayorga Pe\~na, Dami\'an and Mishra, Challenger and Tan, Justin},
    title = {Precision string phenomenology},
    eprint = {2407.13836},
    archivePrefix = {arXiv},
    primaryClass = {hep-th},
    doi = {10.1103/PhysRevD.111.086007},
    journal = {Phys. Rev. D},
    volume = {111},
    number = {8},
    pages = {086007},
    year = {2025}
}

@article{Constantin:2024yxh,
    author = {Constantin, Andrei and Fraser-Taliente, Cristofero S. and Harvey, Thomas R. and Lukas, Andre and Ovrut, Burt},
    title = {Computation of quark masses from string theory},
    eprint = {2402.01615},
    archivePrefix = {arXiv},
    primaryClass = {hep-th},
    doi = {10.1016/j.nuclphysb.2024.116778},
    journal = {Nucl. Phys. B},
    volume = {1010},
    pages = {116778},
    year = {2025}
}

@article{Constantin:2024yaz,
    author = {Constantin, Andrei and Fraser-Taliente, Cristofero S. and Harvey, Thomas R. and Leung, Lucas T.Y. and Lukas, Andre},
    title = {Quark masses and mixing in string-inspired models},
    eprint = {2410.17704},
    archivePrefix = {arXiv},
    primaryClass = {hep-th},
    doi = {10.1007/JHEP06(2025)175},
    journal = {JHEP},
    volume = {06},
    pages = {175},
    year = {2025}
}

@article{Constantin:2025vyt,
    author = {Constantin, Andrei and Leung, Lucas T.-Y. and Lukas, Andre and Nutricati, Luca A.},
    title = {Reproducing {S}tandard {M}odel fermion masses and mixing in string theory: A heterotic line bundle study},
    eprint = {2507.03076},
    archivePrefix = {arXiv},
    primaryClass = {hep-th},
    doi = {10.1103/fbdz-t73z},
    journal = {Phys. Rev. D},
    volume = {113},
    number = {4},
    pages = {046005},
    year = {2026}
}

@article{Mishra:2025xkr,
    author = {Mishra, Challenger and Tan, Justin},
    title = {{H}ermitian {Y}ang--{M}ills connections on general vector bundles: geometry and physical {Y}ukawa couplings},
    eprint = {2512.10907},
    archivePrefix = {arXiv},
    primaryClass = {hep-th},
    year = {2025}
}

@article{Constantin:2026bky,
    author = {Constantin, Andrei and Lukas, Andre and Nutricati, Luca A.},
    title = {{C}alabi--{Y}au Metrics with {K}{\"a}hler Moduli Dependence},
    eprint = {2603.12384},
    archivePrefix = {arXiv},
    primaryClass = {hep-th},
    year = {2026}
}

@article{Kachru:2018lst,
    author = {Kachru, Shamit and Tripathy, Arnav and Zimet, Max},
    title = {{K3} metrics from little string theory},
    eprint = {1810.10540},
    archivePrefix = {arXiv},
    primaryClass = {hep-th},
    year = {2018}
}

@article{Kachru:2020ktm,
    author = {Kachru, Shamit and Tripathy, Arnav and Zimet, Max},
    title = {{K3} metrics},
    eprint = {2006.02435},
    archivePrefix = {arXiv},
    primaryClass = {hep-th},
    year = {2020}
}

@article{Mirjanic:2024srm,
    author = {Mirjani{\'c}, Viktor and Mishra, Challenger},
    title = {Symbolic Approximations to {R}icci-flat Metrics Via Extrinsic Symmetries of {C}alabi--{Y}au Hypersurfaces},
    eprint = {2412.19778},
    archivePrefix = {arXiv},
    primaryClass = {hep-th},
    year = {2025}
}

@article{Li:2020syz,
    author = {Li, Yang},
    title = {Metric {SYZ} conjecture and non-archimedean geometry},
    eprint = {2007.01384},
    archivePrefix = {arXiv},
    primaryClass = {math.DG},
    year = {2020}
}

@article{Braun:2005nv,
    author = {Braun, Volker and He, Yang-Hui and Ovrut, Burt A. and Pantev, Tony},
    title = {The Exact {MSSM} spectrum from string theory},
    eprint = {hep-th/0512177},
    archivePrefix = {arXiv},
    reportNumber = {UPR-1141-T},
    doi = {10.1088/1126-6708/2006/05/043},
    journal = {JHEP},
    volume = {05},
    pages = {043},
    year = {2006}
}

@article{Bouchard:2005ag,
    author = {Bouchard, Vincent and Donagi, Ron},
    title = {An ${SU}(5)$ heterotic standard model},
    eprint = {hep-th/0512149},
    archivePrefix = {arXiv},
    doi = {10.1016/j.physletb.2005.12.042},
    journal = {Phys. Lett. B},
    volume = {633},
    pages = {783--791},
    year = {2006}
}

@article{Bonetti:2016dqh,
    author = {Bonetti, Federico and Weissenbacher, Matthias},
    title = {The {E}uler characteristic correction to the {K}{\"a}hler potential --- revisited},
    eprint = {1608.01300},
    archivePrefix = {arXiv},
    primaryClass = {hep-th},
    reportNumber = {IPMU16-0115, YITP-SB-16-32},
    doi = {10.1007/JHEP01(2017)003},
    journal = {JHEP},
    volume = {01},
    pages = {003},
    year = {2017}
}

@article{Fraser-Taliente:2024etl,
    author = {Fraser-Taliente, Cristofero S. and Harvey, Thomas R. and Kim, Manki},
    title = {Not So Flat Metrics},
    eprint = {2411.00962},
    archivePrefix = {arXiv},
    primaryClass = {hep-th},
    year = {2024},
    month = {11}
}

@article{Nemeschansky:1986yx,
    author = {Nemeschansky, Dennis and Sen, Ashoke},
    title = {Conformal Invariance of Supersymmetric $\sigma$ Models on {C}alabi--{Y}au Manifolds},
    reportNumber = {SLAC-PUB-3925},
    doi = {10.1016/0370-2693(86)91394-8},
    journal = {Phys. Lett. B},
    volume = {178},
    pages = {365--369},
    year = {1986}
}

@article{Becker:2015wga,
    author = {Becker, Katrin and Becker, Melanie and Robbins, Daniel},
    title = {String Corrected Spacetimes and ${SU(N)}$-Structure Manifolds},
    eprint = {1503.04237},
    archivePrefix = {arXiv},
    primaryClass = {hep-th},
    reportNumber = {MIFPA-14-38},
    doi = {10.1016/j.nuclphysb.2015.04.012},
    journal = {Nucl. Phys. B},
    volume = {898},
    pages = {715--735},
    year = {2015}
}

@incollection{Brignole:1997dp,
    author = {Brignole, Andrea and Ib\'{a}\~nez, Luis E. and Mu\~noz, Carlos},
    title = {Soft supersymmetry-breaking terms from supergravity and superstring models},
    booktitle = {Perspectives on Supersymmetry},
    eprint = {hep-ph/9707209},
    archivePrefix = {arXiv},
    primaryClass = {hep-ph},
    doi = {10.1142/9789812839657_0003},
    volume = {18},
    pages = {125--148},
    year = {1998},
    publisher = {World Scientific},
    series = {Advanced Series on Directions in High Energy Physics}
}

@article{Farquet:2012sgm,
    author = {Farquet, Daniel and Scrucca, Claudio A.},
    title = {Scalar geometry and masses in {C}alabi--{Y}au string models},
    eprint = {1205.5728},
    archivePrefix = {arXiv},
    primaryClass = {hep-th},
    doi = {10.1007/JHEP09(2012)025},
    journal = {JHEP},
    volume = {09},
    pages = {025},
    year = {2012}
}

@article{Berg:2010ha,
    author = {Berg, Marcus and Marsh, David and McAllister, Liam and Pajer, Enrico},
    title = {Sequestering in string compactifications},
    eprint = {1012.1858},
    archivePrefix = {arXiv},
    primaryClass = {hep-th},
    doi = {10.1007/JHEP06(2011)134},
    journal = {JHEP},
    volume = {06},
    pages = {134},
    year = {2011}
}

@article{Blesneag:2018mfm,
    author = {Blesneag, {\c S}tefan and Buchbinder, Evgeny I. and Constantin, Andrei and Lukas, Andre and Palti, Eran},
    title = {Matter field K{\"a}hler metric in heterotic string theory from localisation},
    eprint = {1801.09645},
    archivePrefix = {arXiv},
    primaryClass = {hep-th},
    reportNumber = {MPP-2018-5, UUITP-04/18},
    doi = {10.1007/JHEP04(2018)139},
    journal = {JHEP},
    volume = {04},
    pages = {139},
    year = {2018}
}

@article{Anguelova:2010ed,
    author = {Anguelova, Lilia and Quigley, Callum and Sethi, Savdeep},
    title = {The leading quantum corrections to stringy {K}{\"a}hler potentials},
    eprint = {1007.4793},
    archivePrefix = {arXiv},
    primaryClass = {hep-th},
    doi = {10.1007/JHEP10(2010)065},
    journal = {JHEP},
    volume = {10},
    pages = {065},
    year = {2010}
}

@article{Klebanov:2000hb,
    author = {Klebanov, Igor R. and Strassler, Matthew J.},
    title = {Supergravity and a Confining Gauge Theory: Duality Cascades and {$\chi$SB}-Resolution of Naked Singularities},
    eprint = {hep-th/0007191},
    archivePrefix = {arXiv},
    primaryClass = {hep-th},
    doi = {10.1088/1126-6708/2000/08/052},
    journal = {JHEP},
    volume = {08},
    pages = {052},
    year = {2000}
}

@article{Kachru:2002gs,
    author = {Kachru, Shamit and Pearson, John and Verlinde, Herman},
    title = {Brane/Flux Annihilation and the String Dual of a Non-Supersymmetric Field Theory},
    eprint = {hep-th/0112197},
    archivePrefix = {arXiv},
    primaryClass = {hep-th},
    doi = {10.1088/1126-6708/2002/06/021},
    journal = {JHEP},
    volume = {06},
    pages = {021},
    year = {2002}
}

@article{Hebecker:2006jc,
    author = {Hebecker, Arthur and March-Russell, John},
    title = {The Ubiquitous Throat},
    eprint = {hep-th/0607120},
    archivePrefix = {arXiv},
    primaryClass = {hep-th},
    doi = {10.1016/j.nuclphysb.2007.05.003},
    journal = {Nucl. Phys. B},
    volume = {781},
    pages = {99--111},
    year = {2007}
}

@article{Baumann:2010sx,
    author = {Baumann, Daniel and Dymarsky, Anatoly and Kachru, Shamit and Klebanov, Igor R. and McAllister, Liam},
    title = {{D3}-brane Potentials from Fluxes in {AdS/CFT}},
    eprint = {1001.5028},
    archivePrefix = {arXiv},
    primaryClass = {hep-th},
    doi = {10.1007/JHEP06(2010)072},
    journal = {JHEP},
    volume = {06},
    pages = {072},
    year = {2010}
}

@article{Gandhi:2011id,
    author = {Gandhi, Sohang and McAllister, Liam and Sjors, Stefan},
    title = {A Toolkit for Perturbing Flux Compactifications},
    eprint = {1106.0002},
    archivePrefix = {arXiv},
    primaryClass = {hep-th},
    doi = {10.1007/JHEP12(2011)053},
    journal = {JHEP},
    volume = {12},
    pages = {053},
    year = {2011}
}

@article{Anderson:2010mh,
    author = {Anderson, Lara B. and Gray, James and Lukas, Andre and Ovrut, Burt},
    title = {{Stabilizing the Complex Structure in Heterotic Calabi-Yau Vacua}},
    eprint = {1010.0255},
    archivePrefix = {arXiv},
    primaryClass = {hep-th},
    doi = {10.1007/JHEP02(2011)088},
    journal = {JHEP},
    volume = {02},
    pages = {088},
    year = {2011}
}

@article{Anderson:2011cza,
    author = {Anderson, Lara B. and Gray, James and Lukas, Andre and Ovrut, Burt},
    title = {Stabilizing All Geometric Moduli in Heterotic {C}alabi--{Y}au Vacua},
    eprint = {1102.0011},
    archivePrefix = {arXiv},
    primaryClass = {hep-th},
    doi = {10.1103/PhysRevD.83.106011},
    journal = {Phys. Rev. D},
    volume = {83},
    pages = {106011},
    year = {2011}
}

@article{Strominger:1986uh,
    author = {Strominger, Andrew},
    title = {Superstrings with Torsion},
    doi = {10.1016/0550-3213(86)90286-5},
    journal = {Nucl. Phys. B},
    volume = {274},
    pages = {253--284},
    year = {1986}
}

@incollection{McAllister:2023vgy,
	address = {Singapore},
	archiveprefix = {arXiv},
	author = {McAllister, Liam and Quevedo, Fernando},
	booktitle = {Handbook of Quantum Gravity},
	doi = {10.1007/978-981-19-3079-9_58-1},
	editor = {Bambi, Cosimo and Modesto, Leonardo and Shapiro, Ilya},
	eprint = {2310.20559},
	month = {Oct.},
	pages = {1--98},
	primaryclass = {hep-th},
	publisher = {Springer Nature Singapore},
	title = {Moduli Stabilization in String Theory},
	year = {2023}
}

@article{Douglas:2006rr,
    author = {Douglas, Michael R. and Karp, Robert L. and Lukic, Sergio and Reinbacher, Rene},
    title = {Numerical {C}alabi--{Y}au metrics},
    eprint = {hep-th/0612075},
    archivePrefix = {arXiv},
    doi = {10.1063/1.2888403},
    journal = {J. Math. Phys.},
    volume = {49},
    pages = {032302},
    year = {2008}
}

@article{Braun:2007sn,
    author = {Braun, Volker and Brelidze, Tamaz and Douglas, Michael R. and Ovrut, Burt A.},
    title = {{C}alabi--{Y}au Metrics for Quotients and Complete Intersections},
    eprint = {0712.3563},
    archivePrefix = {arXiv},
    primaryClass = {hep-th},
    doi = {10.1088/1126-6708/2008/05/080},
    journal = {JHEP},
    volume = {05},
    pages = {080},
    year = {2008}
}

@article{Anderson:2020hux,
    author = {Anderson, Lara B. and Gerdes, Mathis and Gray, James and Krippendorf, Sven and Raghuram, Nikhil and Ruehle, Fabian},
    title = {Moduli-dependent {C}alabi--{Y}au and ${SU}(3)$-structure metrics from Machine Learning},
    eprint = {2012.04656},
    archivePrefix = {arXiv},
    primaryClass = {hep-th},
    doi = {10.1007/JHEP05(2021)013},
    journal = {JHEP},
    volume = {05},
    pages = {013},
    year = {2021}
}

@article{Witten:1999eg,
    author = {Witten, Edward},
    title = {World-sheet corrections via {D}-instantons},
    eprint = {hep-th/9907041},
    archivePrefix = {arXiv},
    doi = {10.1088/1126-6708/2000/02/030},
    journal = {JHEP},
    volume = {02},
    pages = {030},
    year = {2000}
}

@article{Beasley:2003fx,
    author = {Beasley, Chris and Witten, Edward},
    title = {Residues and world sheet instantons},
    eprint = {hep-th/0304115},
    archivePrefix = {arXiv},
    doi = {10.1088/1126-6708/2003/10/065},
    journal = {JHEP},
    volume = {10},
    pages = {065},
    year = {2003}
}

@article{Gukov:1999ya,
    author = {Gukov, Sergei and Vafa, Cumrun and Witten, Edward},
    title = {{CFT}'s from {C}alabi--{Y}au four folds},
    eprint = {hep-th/9906070},
    archivePrefix = {arXiv},
    doi = {10.1016/S0550-3213(00)00373-4},
    journal = {Nucl. Phys. B},
    volume = {584},
    pages = {69--108},
    year = {2000},
    note = {[Erratum: Nucl. Phys. B 608, 477--478 (2001)]}
}

@article{Giddings:2001yu,
    author = {Giddings, Steven B. and Kachru, Shamit and Polchinski, Joseph},
    title = {Hierarchies from Fluxes in String Compactifications},
    eprint = {hep-th/0105097},
    archivePrefix = {arXiv},
    doi = {10.1103/PhysRevD.66.106006},
    journal = {Phys. Rev. D},
    volume = {66},
    pages = {106006},
    year = {2002}
}

@article{Kachru:2003aw,
    author = {Kachru, Shamit and Kallosh, Renata and Linde, Andrei D. and Trivedi, Sandip P.},
    title = {De~{S}itter Vacua in String Theory},
    eprint = {hep-th/0301240},
    archivePrefix = {arXiv},
    doi = {10.1103/PhysRevD.68.046005},
    journal = {Phys. Rev. D},
    volume = {68},
    pages = {046005},
    year = {2003}
}

@article{Balasubramanian:2005zx,
    author = {Balasubramanian, Vijay and Berglund, Per and Conlon, Joseph P. and Quevedo, Fernando},
    title = {Systematics of Moduli Stabilisation in {C}alabi--{Y}au Flux Compactifications},
    eprint = {hep-th/0502058},
    archivePrefix = {arXiv},
    doi = {10.1088/1126-6708/2005/03/007},
    journal = {JHEP},
    volume = {03},
    pages = {007},
    year = {2005}
}

@article{Svrcek:2006yi,
    author = {Svrcek, Peter and Witten, Edward},
    title = {Axions In String Theory},
    eprint = {hep-th/0605206},
    archivePrefix = {arXiv},
    doi = {10.1088/1126-6708/2006/06/051},
    journal = {JHEP},
    volume = {06},
    pages = {051},
    year = {2006}
}

@article{Arvanitaki:2009fg,
    author = {Arvanitaki, Asimina and Dimopoulos, Savas and Dubovsky, Sergei and Kaloper, Nemanja and March-Russell, John},
    title = {String Axiverse},
    eprint = {0905.4720},
    archivePrefix = {arXiv},
    primaryClass = {hep-th},
    doi = {10.1103/PhysRevD.81.123530},
    journal = {Phys. Rev. D},
    volume = {81},
    pages = {123530},
    year = {2010}
}

@article{Marsh:2015xka,
    author = {Marsh, David J. E.},
    title = {Axion Cosmology},
    eprint = {1510.07633},
    archivePrefix = {arXiv},
    primaryClass = {astro-ph.CO},
    doi = {10.1016/j.physrep.2016.06.005},
    journal = {Phys. Rept.},
    volume = {643},
    pages = {1--79},
    year = {2016}
}

@article{Dimopoulos:2005ac,
    author = {Dimopoulos, Savas and Kachru, Shamit and McGreevy, John and Wacker, Jay G.},
    title = {{N}-flation},
    eprint = {hep-th/0507205},
    archivePrefix = {arXiv},
    doi = {10.1088/1475-7516/2008/08/003},
    journal = {JCAP},
    volume = {08},
    pages = {003},
    year = {2008}
}

@article{Kim:2004rp,
    author = {Kim, Jihn E. and Nilles, Hans Peter and Peloso, Marco},
    title = {Completing natural inflation},
    eprint = {hep-ph/0409138},
    archivePrefix = {arXiv},
    doi = {10.1088/1475-7516/2005/01/005},
    journal = {JCAP},
    volume = {01},
    pages = {005},
    year = {2005}
}

@article{Cicoli:2012sz,
    author = {Cicoli, Michele and Goodsell, Mark D. and Ringwald, Andreas},
    title = {The type {IIB} string axiverse and its low-energy phenomenology},
    eprint = {1206.0819},
    archivePrefix = {arXiv},
    primaryClass = {hep-th},
    doi = {10.1007/JHEP10(2012)146},
    journal = {JHEP},
    volume = {10},
    pages = {146},
    year = {2012}
}

@article{Demirtas:2018akl,
    author = {Demirtas, Mehmet and Long, Cody and McAllister, Liam and Stillman, Mike},
    title = {The {K}reuzer--{S}karke Axiverse},
    eprint = {1808.01282},
    archivePrefix = {arXiv},
    primaryClass = {hep-th},
    doi = {10.1007/JHEP04(2020)138},
    journal = {JHEP},
    volume = {04},
    pages = {138},
    year = {2020}
}

@article{Arkani-Hamed:2006emk,
    author = {Arkani-Hamed, Nima and Motl, Lubos and Nicolis, Alberto and Vafa, Cumrun},
    title = {The String landscape, black holes and gravity as the weakest force},
    eprint = {hep-th/0601001},
    archivePrefix = {arXiv},
    doi = {10.1088/1126-6708/2007/06/060},
    journal = {JHEP},
    volume = {06},
    pages = {060},
    year = {2007}
}

@article{Ooguri:2006in,
    author = {Ooguri, Hirosi and Vafa, Cumrun},
    title = {On the Geometry of the String Landscape and the Swampland},
    eprint = {hep-th/0605264},
    archivePrefix = {arXiv},
    doi = {10.1016/j.nuclphysb.2006.10.033},
    journal = {Nucl. Phys. B},
    volume = {766},
    pages = {21--33},
    year = {2007}
}

@article{Palti:2019pca,
    author = {Palti, Eran},
    title = {The Swampland: Introduction and Review},
    eprint = {1903.06239},
    archivePrefix = {arXiv},
    primaryClass = {hep-th},
    doi = {10.1002/prop.201900037},
    journal = {Fortsch. Phys.},
    volume = {67},
    number = {6},
    pages = {1900037},
    year = {2019}
}

@article{Ashmore:2019wzb,
    author = {Ashmore, Anthony and He, Yang-Hui and Ovrut, Burt A.},
    title = {Machine Learning {C}alabi--{Y}au Metrics},
    eprint = {1910.08605},
    archivePrefix = {arXiv},
    primaryClass = {hep-th},
    doi = {10.1002/prop.202000068},
    journal = {Fortsch. Phys.},
    volume = {68},
    number = {9},
    pages = {2000068},
    year = {2020}
}

@article{Ashmore:2021qcb,
    author = {Ashmore, Anthony and Calmon, Lucille and He, Yang-Hui and Ovrut, Burt A.},
    title = {{C}alabi--{Y}au Metrics, Energy Functionals and Machine-Learning},
    eprint = {2112.10872},
    archivePrefix = {arXiv},
    primaryClass = {hep-th},
    doi = {10.1142/S2810939222500034},
    journal = {Int. J. Data Sci. Math. Sci.},
    volume = {1},
    number = {1},
    pages = {49--61},
    year = {2023}
}

@article{Ashmore:2021qdf,
    author = {Ashmore, Anthony and Ruehle, Fabian},
    title = {Moduli-dependent {KK} towers and the swampland distance conjecture on the quintic {C}alabi--{Y}au manifold},
    eprint = {2103.07472},
    archivePrefix = {arXiv},
    primaryClass = {hep-th},
    reportNumber = {CERN-TH-2021-032},
    doi = {10.1103/PhysRevD.103.106028},
    journal = {Phys. Rev. D},
    volume = {103},
    number = {10},
    pages = {106028},
    year = {2021}
}

@article{Grimm:2019ixq,
    author = {Grimm, Thomas W. and Ruehle, Fabian and van de Heisteeg, Damian},
    title = {Classifying {C}alabi--{Y}au Threefolds Using Infinite Distance Limits},
    eprint = {1910.02963},
    archivePrefix = {arXiv},
    primaryClass = {hep-th},
    doi = {10.1007/s00220-021-03972-9},
    journal = {Commun. Math. Phys.},
    volume = {382},
    number = {1},
    pages = {239--275},
    year = {2021}
}

@article{Brodie:2021toe,
    author = {Brodie, Callum R. and Constantin, Andrei and Lukas, Andre and Ruehle, Fabian},
    title = {Swampland Conjectures and Infinite Flop Chains},
    eprint = {2104.03325},
    archivePrefix = {arXiv},
    primaryClass = {hep-th},
    reportNumber = {CERN-TH-2021-051},
    doi = {10.1103/PhysRevD.104.046008},
    journal = {Phys. Rev. D},
    volume = {104},
    number = {4},
    pages = {046008},
    year = {2021}
}

@article{Brodie:2021xsf,
    author = {Brodie, Callum R. and Constantin, Andrei and Lukas, Andre and Ruehle, Fabian},
    title = {Geodesics in the Extended {K}{\"a}hler Cone of {C}alabi--{Y}au Threefolds},
    eprint = {2108.10323},
    archivePrefix = {arXiv},
    primaryClass = {hep-th},
    reportNumber = {CERN-TH-2021-123},
    doi = {10.1007/JHEP03(2022)024},
    journal = {JHEP},
    volume = {03},
    pages = {024},
    year = {2022}
}

@article{Larfors:2022nep,
    author = {Larfors, Magdalena and Lukas, Andre and Ruehle, Fabian and Schneider, Robin},
    title = {Numerical Metrics for Complete Intersection and {K}reuzer--{S}karke {C}alabi--{Y}au Manifolds},
    eprint = {2205.13408},
    archivePrefix = {arXiv},
    primaryClass = {hep-th},
    reportNumber = {UUITP-25/22},
    doi = {10.1088/2632-2153/ac8e4e},
    journal = {Mach. Learn. Sci. Tech.},
    volume = {3},
    number = {3},
    pages = {035014},
    year = {2022}
}

@article{Heidenreich:2018kpg,
    author = {Heidenreich, Ben and Reece, Matthew and Rudelius, Tom},
    title = {Emergence of Weak Coupling at Large Distance in Quantum Gravity},
    eprint = {1802.08698},
    archivePrefix = {arXiv},
    primaryClass = {hep-th},
    doi = {10.1103/PhysRevLett.121.051601},
    journal = {Phys. Rev. Lett.},
    volume = {121},
    number = {5},
    pages = {051601},
    year = {2018}
}

@article{MurgasIbarra:2024lks,
    author = {Murgas Ibarra, Javier Jos{\'e} and Oehlmann, Paul-Konstantin and Ruehle, Fabian and Svanes, Eirik Eik},
    title = {A Heterotic {K}{\"a}hler Gravity and the Distance Conjecture},
    eprint = {2406.04393},
    archivePrefix = {arXiv},
    primaryClass = {hep-th},
    doi = {10.1007/JHEP01(2025)168},
    journal = {JHEP},
    volume = {01},
    pages = {168},
    year = {2025}
}

@article{Ahmed:2023cnw,
    author = {Ahmed, Hamza and Ruehle, Fabian},
    title = {Level crossings, attractor points and complex multiplication},
    eprint = {2304.00027},
    archivePrefix = {arXiv},
    primaryClass = {hep-th},
    doi = {10.1007/JHEP06(2023)164},
    journal = {JHEP},
    volume = {06},
    pages = {164},
    year = {2023}
}

@article{Harvey:1982xk,
    author = {Harvey, Reese and Lawson, Jr., H. Blaine},
    title = {Calibrated geometries},
    doi = {10.1007/BF02392726},
    journal = {Acta Math.},
    volume = {148},
    pages = {47--157},
    year = {1982}
}

@article{McLean:1998ub,
    author = {McLean, Robert C.},
    title = {Deformations of calibrated submanifolds},
    doi = {10.4310/CAG.1998.v6.n4.a4},
    journal = {Commun. Anal. Geom.},
    volume = {6},
    number = {4},
    pages = {705--747},
    year = {1998}
}

@article{Candelas:1990rm,
    author = {Candelas, Philip and de la Ossa, Xenia C. and Green, Paul S. and Parkes, Linda},
    title = {A Pair of {C}alabi--{Y}au manifolds as an exactly soluble superconformal theory},
    doi = {10.1016/0550-3213(91)90292-6},
    journal = {Nucl. Phys. B},
    volume = {359},
    pages = {21--74},
    year = {1991}
}

@inproceedings{Kontsevich:2000hb,
    author = {Kontsevich, Maxim and Soibelman, Yan},
    title = {Homological mirror symmetry and torus fibrations},
    booktitle = {Symplectic Geometry and Mirror Symmetry},
    eprint = {math/0011041},
    archivePrefix = {arXiv},
    primaryClass = {math.SG},
    pages = {203--263},
    year = {2001}
}

@article{Auroux:2009hf,
    author = {Auroux, Denis},
    title = {Special Lagrangian fibrations, wall-crossing, and mirror symmetry},
    eprint = {0902.1595},
    archivePrefix = {arXiv},
    primaryClass = {math.SG},
    journal = {Surveys Diff. Geom.},
    volume = {13},
    pages = {1--47},
    year = {2009}
}

@article{Gross:2007pf,
    author = {Gross, Mark and Siebert, Bernd},
    title = {From real affine geometry to complex geometry},
    eprint = {math/0703822},
    archivePrefix = {arXiv},
    primaryClass = {math.AG},
    doi = {10.4007/annals.2011.174.3.1},
    journal = {Annals Math.},
    volume = {174},
    pages = {1301--1428},
    year = {2011}
}

@article{Abouzaid:2012kk,
    author = {Abouzaid, Mohammed and Auroux, Denis and Katzarkov, Ludmil},
    title = {Lagrangian fibrations on blowups of toric varieties and mirror symmetry for hypersurfaces},
    eprint = {1205.0053},
    archivePrefix = {arXiv},
    primaryClass = {math.SG},
    doi = {10.1007/s10240-016-0081-9},
    journal = {Publ. Math. IHES},
    volume = {123},
    pages = {199--282},
    year = {2016}
}

@article{Douglas:2024pmn,
    author = "Douglas, Michael R. and Platt, Daniel and Qi, Yidi and Barbosa, Rodrigo",
    title = "{Harmonic $1$-forms on real loci of Calabi-Yau manifolds}",
    eprint = "2405.19402",
    archivePrefix = "arXiv",
    primaryClass = "math.DG",
    month = "5",
    year = "2024"
}

@article{Qi2020,
    author = {Douglas, Michael R. and Lakshminarasimhan, Subramanian and Qi, Yidi},
    title = {{MLG}eometry},
    journal = {\!},
    pages = {\url{https://github.com/yidiq7/MLGeometry}},
    year = {2020}
}

@article{Reid:1987,
    author = "Reid, Miles",
    title = "{The moduli space of 3-folds with K=0 may nevertheless be irreducible}",
    doi = "10.1007/BF01458074",
    journal = "Math. Ann.",
    volume = "278",
    pages = "329--334",
    year = "1987"
}

@article{Candelas:1990pi,
    author = "Candelas, Philip and de la Ossa, Xenia",
    title = "{Moduli Space of {Calabi-Yau} Manifolds}",
    reportNumber = "UTTG-07-90",
    doi = "10.1016/0550-3213(91)90122-E",
    journal = "Nucl. Phys. B",
    volume = "355",
    pages = "455--481",
    year = "1991"
}

@article{Candelas:1985en,
    author = "Candelas, Philip and Horowitz, Gary T. and Strominger, Andrew and Witten, Edward",
    title = "{Vacuum Configurations for Superstrings}",
    doi = "10.1016/0550-3213(85)90602-9",
    journal = "Nucl. Phys. B",
    volume = "258",
    pages = "46--74",
    year = "1985"
}

@article{Strominger:1985it,
    author = "Strominger, Andrew",
    title = "{Yukawa Couplings in Superstring Compactification}",
    doi = "10.1103/PhysRevLett.55.2547",
    journal = "Phys. Rev. Lett.",
    volume = "55",
    pages = "2547",
    year = "1985"
}

@article{Donaldson:1985zz,
    author = "Donaldson, S. K.",
    title = "{Anti self-dual Yang-Mills connections over complex algebraic surfaces and stable vector bundles}",
    doi = "10.1112/plms/s3-50.1.1",
    journal = "Proc. Lond. Math. Soc.",
    volume = "50",
    pages = "1--26",
    year = "1985"
}

@article{Uhlenbeck:1986de,
    author = "Uhlenbeck, K. and Yau, S. T.",
    title = "{On the existence of Hermitian-Yang-Mills connections in stable vector bundles}",
    doi = "10.1002/cpa.3160390714",
    journal = "Commun. Pure Appl. Math.",
    volume = "39",
    number = "S1",
    pages = "S257--S293",
    year = "1986"
}

@inproceedings{Li:1986hh,
    author = "Li, Jun and Yau, Shing-Tung",
    title = "{Hermitian-Yang-Mills Connection on non-Kahler Manifolds}",
    booktitle = "{Mathematical Aspects of String Theory}",
    pages = "560--573",
    publisher = "World Scientific",
    address = "Singapore",
    year = "1987"
}

@article{Hull:1986kz,
    author = "Hull, C. M.",
    title = "{Compactifications of the Heterotic Superstring}",
    doi = "10.1016/0370-2693(86)91393-6",
    journal = "Phys. Lett. B",
    volume = "178",
    pages = "357--364",
    year = "1986"
}

@article{Huang:2018esr,
    author = "Huang, Yi-Chun and Taylor, Washington",
    title = "{On the prevalence of elliptic and genus one fibrations among toric hypersurface Calabi-Yau threefolds}",
    eprint = "1809.05160",
    archivePrefix = "arXiv",
    primaryClass = "hep-th",
    doi = "10.1007/JHEP03(2019)014",
    journal = "JHEP",
    volume = "03",
    pages = "014",
    year = "2019"
}

@article{Goldstein:2002pg,
    author = "Goldstein, Edward and Prokushkin, Sergey",
    title = "{Geometric model for complex nonKahler manifolds with SU(3) structure}",
    eprint = "hep-th/0212307",
    archivePrefix = "arXiv",
    doi = "10.1007/s00220-004-1177-7",
    journal = "Commun. Math. Phys.",
    volume = "251",
    pages = "65--78",
    year = "2004"
}

@article{Fu:2006vj,
    author = "Fu, Ji-Xiang and Yau, Shing-Tung",
    title = "{The theory of superstring with flux on non-Kahler manifolds and the complex Monge-Ampere equation}",
    eprint = "hep-th/0604063",
    archivePrefix = "arXiv",
    doi = "10.4310/jdg/1214454218",
    journal = "J. Diff. Geom.",
    volume = "78",
    pages = "369--428",
    year = "2008"
}

@article{Danielsson:2018ztv,
    author = "Danielsson, Ulf H. and Van Riet, Thomas",
    title = "{What if string theory has no de Sitter vacua?}",
    eprint = "1804.01120",
    archivePrefix = "arXiv",
    primaryClass = "hep-th",
    doi = "10.1142/S0218271818300070",
    journal = "Int. J. Mod. Phys. D",
    volume = "27",
    number = "12",
    pages = "1830007",
    year = "2018"
}

@article{Bena:2011wh,
    author = "Bena, Iosif and Giecold, Gregory and Grana, Mariana and Halmagyi, Nick and Massai, Stefano",
    title = "{The backreaction of anti-D3 branes on the Klebanov-Strassler geometry}",
    eprint = "1106.6165",
    archivePrefix = "arXiv",
    primaryClass = "hep-th",
    doi = "10.1007/JHEP06(2013)060",
    journal = "JHEP",
    volume = "06",
    pages = "060",
    year = "2013"
}

@article{Avram:1996xn,
    author = "Avram, A. C. and Kreuzer, M. and Mandelberg, M. and Skarke, H.",
    title = "{Searching for K3 fibrations}",
    eprint = "hep-th/9610154",
    archivePrefix = "arXiv",
    doi = "10.1016/S0550-3213(97)00214-9",
    journal = "Nucl. Phys. B",
    volume = "494",
    pages = "567--589",
    year = "1997"
}

@article{Candelas:2012az,
    author = "Candelas, Philip and Constantin, Andrei and Skarke, Harald",
    title = "{An Abundance of K3 Fibrations from Polyhedra with Interchangeable Parts}",
    eprint = "1207.4792",
    archivePrefix = "arXiv",
    primaryClass = "hep-th",
    doi = "10.1007/s00220-013-1802-2",
    journal = "Commun. Math. Phys.",
    volume = "324",
    pages = "937--959",
    year = "2013"
}

@article{Anderson:2017aux,
    author = "Anderson, Lara B. and Gao, Xin and Gray, James and Lee, Seung-Joo",
    title = "{Fibrations in CICY Threefolds}",
    eprint = "1708.07907",
    archivePrefix = "arXiv",
    primaryClass = "hep-th",
    doi = "10.1007/JHEP10(2017)077",
    journal = "JHEP",
    volume = "10",
    pages = "077",
    year = "2017"
}

@inproceedings{Doran:2016uew,
    author = "Doran, Charles F. and Harder, Andrew and Thompson, Alan",
    title = "{Mirror symmetry, Tyurin degenerations and fibrations on Calabi-Yau manifolds}",
    eprint = "1601.08110",
    archivePrefix = "arXiv",
    primaryClass = "math.AG",
    booktitle = "{String-Math 2015}",
    series = "Proc. Symp. Pure Math.",
    volume = "96",
    pages = "93--131",
    year = "2017"
}

@article{Kanazawa:2016tnt,
    author = "Kanazawa, Atsushi",
    title = "{Doran--Harder--Thompson Conjecture via SYZ Mirror Symmetry: Elliptic Curves}",
    eprint = "1612.04623",
    archivePrefix = "arXiv",
    primaryClass = "math.AG",
    doi = "10.3842/SIGMA.2017.024",
    journal = "SIGMA",
    volume = "13",
    pages = "024",
    year = "2017"
}

@article{Boucksom:2016khs,
    author = "Boucksom, Sebastien and Jonsson, Mattias",
    title = "{Tropical and non-Archimedean limits of degenerating families of volume forms}",
    eprint = "1605.05277",
    archivePrefix = "arXiv",
    primaryClass = "math.DG",
    doi = "10.5802/jep.39",
    journal = "J. Ec. polytech. Math.",
    volume = "4",
    pages = "87--139",
    year = "2017"
}

@article{Tosatti:2020nyp,
    author = "Tosatti, Valentino",
    title = "{Collapsing Calabi-Yau manifolds}",
    eprint = "2003.00673",
    archivePrefix = "arXiv",
    primaryClass = "math.DG",
    journal = "Surveys Diff. Geom.",
    volume = "23",
    pages = "305--337",
    year = "2020"
}

@article{Li:2022kqp,
    author = "Li, Yang",
    title = "{Survey on the metric SYZ conjecture and non-Archimedean geometry}",
    eprint = "2204.11363",
    archivePrefix = "arXiv",
    primaryClass = "math.AG",
    doi = "10.1142/S0217751X22300095",
    journal = "Int. J. Mod. Phys. A",
    volume = "37",
    number = "17",
    pages = "2230009",
    year = "2022"
}

@article{Hull:1985ab,
    author = "Hull, C. M.",
    title = "{Anomalies, Ambiguities and Superstrings}",
    doi = "10.1016/0370-2693(86)90544-7",
    journal = "Phys. Lett. B",
    volume = "167",
    pages = "51--55",
    year = "1986"
}

@article{Geroch:1967fs,
    author = "Geroch, Robert",
    title = "{Topology in general relativity}",
    doi = "10.1063/1.1705276",
    journal = "J. Math. Phys.",
    volume = "8",
    pages = "782--786",
    year = "1967"
}

@article{AQR2026,
    author = "Ahmed, Hamza and Qi, Yidi and and Ruehle, Fabian",
    title = "unpublished",
    journal = "to appear",
    year = "2026" 
}

@article{Lee:2025pue,
    author = "Lee, Seung-Joo and Lukas, Andre",
    title = "{Approximate Ricci-flat Metrics for Calabi-Yau Manifolds}",
    eprint = "2506.15766",
    archivePrefix = "arXiv",
    primaryClass = "hep-th",
    month = "6",
    year = "2025"
}

@article{Batyrev:1993oya,
    author = "Batyrev, Victor V.",
    title = "{Dual polyhedra and mirror symmetry for Calabi-Yau hypersurfaces in toric varieties}",
    eprint = "alg-geom/9310003",
    archivePrefix = "arXiv",
    journal = "J. Alg. Geom.",
    volume = "3",
    pages = "493--545",
    year = "1994"
}

@article{Scholler:2018apc,
    author = {Sch{\"o}ller, Friedrich and Skarke, Harald},
    title = "{All Weight Systems for Calabi{\textendash}Yau Fourfolds from Reflexive Polyhedra}",
    eprint = "1808.02422",
    archivePrefix = "arXiv",
    primaryClass = "hep-th",
    doi = "10.1007/s00220-019-03331-9",
    journal = "Commun. Math. Phys.",
    volume = "372",
    number = "2",
    pages = "657--678",
    year = "2019"
}

@article{Heyes:2026rch,
    author = "Heyes, Elli and Hirst, Edward and Earp, Henrique N. S{\'a} and Silva, Tom{\'a}s S. R.",
    title = "{Neural and numerical methods for $\mathrm{G}_2$-structures on contact Calabi-Yau 7-manifolds}",
    eprint = "2602.12438",
    archivePrefix = "arXiv",
    primaryClass = "math.DG",
    month = "2",
    year = "2026"
}

@article{Aggarwal:2023swe,
    author = "Aggarwal, Daattavya and He, Yang-Hui and Heyes, Elli and Hirst, Edward and Earp, Henrique N. S{\'a} and Silva, Tom{\'a}s S. R.",
    title = "{Machine learning Sasakian and G2 topology on contact Calabi-Yau 7-manifolds}",
    eprint = "2310.03064",
    archivePrefix = "arXiv",
    primaryClass = "math.DG",
    reportNumber = "QMUL-PH-23-14",
    doi = "10.1016/j.physletb.2024.138517",
    journal = "Phys. Lett. B",
    volume = "850",
    pages = "138517",
    year = "2024"
}

@article{Hendi:2024yin,
    author = "Hendi, Yacoub and Larfors, Magdalena and Walden, Moritz",
    title = "{Learning group invariant Calabi{\textendash}Yau metrics by fundamental domain projections}",
    eprint = "2407.06914",
    archivePrefix = "arXiv",
    primaryClass = "hep-th",
    reportNumber = "UUITP-21/24",
    doi = "10.1088/2632-2153/adb4bb",
    journal = "Mach. Learn. Sci. Tech.",
    volume = "6",
    number = "1",
    pages = "015050",
    year = "2025"
}

@article{Li:2019oyj,
    author = "Li, Yang",
    title = "{SYZ geometry for Calabi-Yau 3-folds: Taub-NUT and Ooguri-Vafa type metrics}",
    eprint = "1902.08770",
    archivePrefix = "arXiv",
    primaryClass = "math.DG",
    month = "2",
    year = "2019"
}

@article{li2022strominger,
  title={Strominger--Yau--Zaslow conjecture for Calabi--Yau hypersurfaces in the Fermat family},
  author={Li, Yang},
  journal={Acta Mathematica},
  volume={229},
  number={1},
  pages={1--53},
  year={2022},
  publisher={Lehigh University Bethlehem, Penn., USA}
}

@article{Denef:2006ad,
    author = "Denef, Frederik and Douglas, Michael R.",
    title = "{Computational complexity of the landscape. I.}",
    eprint = "hep-th/0602072",
    archivePrefix = "arXiv",
    doi = "10.1016/j.aop.2006.07.013",
    journal = "Annals Phys.",
    volume = "322",
    pages = "1096--1142",
    year = "2007"
}

@article{Denef:2004ze,
    author = "Denef, Frederik and Douglas, Michael R.",
    title = "{Distributions of flux vacua}",
    eprint = "hep-th/0404116",
    archivePrefix = "arXiv",
    doi = "10.1088/1126-6708/2004/05/072",
    journal = "JHEP",
    volume = "05",
    pages = "072",
    year = "2004"
}

@article{Li:2025mjm,
    author = "Li, Yang",
    title = "{Intermediate complex structure limit for Calabi-Yau metrics}",
    doi = "10.1007/s00222-024-01314-9",
    journal = "Invent. Math.",
    volume = "240",
    number = "2",
    pages = "459--496",
    year = "2025"
}

@proceedings{BIRS:2026,
	organization = {{Banff International Research Station for Mathematical Innovation and Discovery} and {Chennai Mathematical Institute}},
	publisher = {\url{https://www.birs.ca/events/2026/5-day-workshops/26w5653}},
	title = {Recent Progress in Computational String Geometry},
	year = {26--31 January 2026}}

@article{Li:2025deg,
    author = "Li, Yang",
    title = "{Degeneration of Calabi-Yau metrics and canonical basis}",
    eprint = "2505.11087",
    archivePrefix = "arXiv",
    primaryClass = "math.DG",
    year = "2025"
}

@article{LiTosatti:2025,
    author = "Li, Yang and Tosatti, Valentino",
    title = "{Generic regularity of intermediate complex structure limits}",
    eprint = "2511.04651",
    archivePrefix = "arXiv",
    primaryClass = "math.DG",
    doi = "10.1515/crelle-2026-0013",
    journal = "J. Reine Angew. Math.",
    year = "2026"
}

@article{Li:2026syz,
    author = "Li, Yang",
    title = "{Valuative independence and metric SYZ conjecture}",
    eprint = "2605.00516",
    archivePrefix = "arXiv",
    primaryClass = "math.DG",
    year = "2026"
}

@article{Blum:2026viy,
    author = "Blum, Harold and Liu, Yuchen",
    title = "{Valuative independence for Calabi--Yau varieties}",
    eprint = "2604.27890",
    archivePrefix = "arXiv",
    primaryClass = "math.AG",
    year = "2026"
}

@article{Shiffman:1998,
    author = "Shiffman, Bernard and Zelditch, Steve",
    title = "{Distribution of zeros of random and quantum chaotic sections of positive line bundles}",
    eprint = "math/9803052",
    archivePrefix = "arXiv",
    primaryClass = "math.CV",
    doi = "10.1007/s002200050544",
    journal = "Commun. Math. Phys.",
    volume = "200",
    pages = "661--683",
    year = "1999"
}

@book{Bertrand:1889,
    author = "Bertrand, Joseph",
    title = "{Calcul des probabilit\'es}",
    publisher = "Gauthier-Villars",
    address = "Paris",
    year = "1889"
}

@article{Jaynes:1973,
    author = "Jaynes, Edwin T.",
    title = "{The Well-Posed Problem}",
    doi = "10.1007/BF00709116",
    journal = "Found. Phys.",
    volume = "3",
    number = "4",
    pages = "477--493",
    year = "1973"
}

@article{Berglund:unpubPointSelection,
    author = "Berglund, Per and Butbaia, Giorgi and H{\"u}bsch, Tristan and Jejjala, Vishnu and Mayorga Pe{\~n}a, Dami{\'a}n and Mishra, Challenger and Tan, Justin",
    title = "{Point Selection for Spectral Network Approximations to Ricci-flat Calabi--Yau Metrics}",
    journal = "Unpublished results presented at String Data, CalTech",
    year = "2023"
}

@article{Keller:2009,
    author = "Keller, Julien and Lukic, Sergio",
    title = "{Numerical Weil-Petersson Metrics on Moduli Spaces of Calabi-Yau Manifolds}",
    eprint = "0907.1387",
    archivePrefix = "arXiv",
    primaryClass = "math.DG",
    doi = "10.1016/j.geomphys.2015.02.018",
    journal = "J. Geom. Phys.",
    volume = "92",
    pages = "252--270",
    year = "2015"
}

@article{Ruehle:Pollica2025,
    author = "Ruehle, Fabian",
    title = "{Backreaction of Fluxes on Calabi--Yau Metrics}",
    journal = "{Talk at the Pollica Workshop ``Calabi--Yau Manifolds'', June 2--6, 2025}",
    year = "2025"
}

@article{Dasgupta:1999ss,
    author = "Dasgupta, Keshav and Rajesh, Govindan and Sethi, Savdeep",
    title = "{M theory, orientifolds and G - flux}",
    eprint = "hep-th/9908088",
    archivePrefix = "arXiv",
    reportNumber = "IASSNS-HEP-99-75, NSF-ITP-99-095",
    doi = "10.1088/1126-6708/1999/08/023",
    journal = "JHEP",
    volume = "08",
    pages = "023",
    year = "1999"
}

@article{Melnikov:2014ywa,
    author = "Melnikov, Ilarion V. and Minasian, Ruben and Sethi, Savdeep",
    title = "{Heterotic fluxes and supersymmetry}",
    eprint = "1403.4298",
    archivePrefix = "arXiv",
    primaryClass = "hep-th",
    reportNumber = "EFI-14-5, IPHT-T14-026, MIFPA-14-07",
    doi = "10.1007/JHEP06(2014)174",
    journal = "JHEP",
    volume = "06",
    pages = "174",
    year = "2014"
}

@article{Sethi:2017phn,
    author = "Sethi, Savdeep",
    title = "{Supersymmetry Breaking by Fluxes}",
    eprint = "1709.03554",
    archivePrefix = "arXiv",
    primaryClass = "hep-th",
    reportNumber = "EFI-17-5",
    doi = "10.1007/JHEP10(2018)022",
    journal = "JHEP",
    volume = "10",
    pages = "022",
    year = "2018"
}

@article{Lust:2026mys,
    author = {L{\"u}st, Severin and Ruehle, Fabian and Schreyer, Simon},
    title = "{Warped Numerical Calabi-Yau Metrics}",
    eprint = "2607.18402",
    archivePrefix = "arXiv",
    primaryClass = "hep-th",
    month = "7",
    year = "2026"
}

\end{document}